\RequirePackage{fix-cm}
\documentclass[epjc3,10pt]{svjour3}

\smartqed  
\RequirePackage{graphicx}

\usepackage{xspace}
\usepackage{amsmath}
\usepackage{amssymb}
\usepackage{paralist}
\graphicspath{ {figures/} }

\newcommand{\includegraphicss}[2][]{{\includegraphics[#1]{#2}}}
\usepackage{footmisc}

\usepackage[mathlines,displaymath]{lineno}
\renewcommand{\arraystretch}{1.25} 

\newcommand{\NNLOJET}{NNLO\protect\scalebox{0.8}{JET}}
\makeatletter
\newcommand*{\rom}[1]{\expandafter\@slowromancap\romannumeral #1@}
\makeatother

\newcommand{\Qsq}{\ensuremath{Q^2}\xspace}
\newcommand{\mjj}{\ensuremath{M_\mathrm{12}}\xspace}

\newcommand{\invpb}{\ensuremath{{\rm pb}^{-1}}\xspace}
\newcommand{\GeV}{\ensuremath{\rm GeV}\xspace}
\newcommand{\GeVsq}{\ensuremath{\GeV^2}\xspace}

\newcommand{\etalab}{\ensuremath{\eta_{\rm lab}^{\rm jet}}\xspace}
\newcommand{\etaavg}{\ensuremath{\langle\eta_{\rm lab}^{\rm jet}\rangle}\xspace}
\newcommand{\deleta}{\ensuremath{\Delta\eta_{\rm lab}^{\rm jet}}\xspace}
\newcommand{\etaj}{\ensuremath{\eta_{\rm jet}^{\ast}}\xspace}

\newcommand{\D}{\mathrm{d}}

\newcommand{\deletastar}{\ensuremath{\Delta\eta^{\ast}}\xspace}

\newcommand{\muf}{\ensuremath{\mu_{\text{F}}}\xspace}
\newcommand{\mur}{\ensuremath{\mu_{\text{R}}}\xspace}
\newcommand{\as}{\ensuremath{\alpha_{\text{s}}}\xspace}
\newcommand{\asmz}{\ensuremath{\as(M_{\text{Z}})}\xspace}
\newcommand{\asmur}{\ensuremath{\as(\mur)}\xspace}
\newcommand{\ptjet}{\ensuremath{p_{\mathrm{T}}^{\ast,\rm jet}}\xspace}
\newcommand{\ptjone}{\ensuremath{p_{\mathrm{T}}^{\ast,\rm jet1}}\xspace}
\newcommand{\ptjtwo}{\ensuremath{p_{\mathrm{T}}^{\ast,\rm jet2}}\xspace}
\newcommand{\ptavg}{\ensuremath{\langle p_{\mathrm{T}}\rangle }\xspace}
\newcommand{\pom}{\ensuremath{I\!\!P}}

\newcommand{\mxq}{\ensuremath{M_{\rm X}^{2}}\xspace}
\newcommand{\mx}{\ensuremath{M_{\rm X}}\xspace}
\newcommand{\MY}{\ensuremath{M_{\rm Y}}\xspace}
\newcommand{\xpom}{\ensuremath{x_{\pom}}\xspace}
\newcommand{\zpom}{\ensuremath{z^{\rm obs}_{\pom}}\xspace}
\newcommand{\zint}{\ensuremath{z_{\pom}}\xspace}
\newcommand{\xbj}{\ensuremath{x_{\mathrm{Bj}}\xspace}}

\newcommand{\ok}{\checkmark}
\newcommand{\chisq}{\ensuremath{\chi^{2}\xspace}}
\newcommand{\ndf}{\ensuremath{n_{\rm dof}\xspace}}

\newcommand{\HERAI} {\protect\scalebox{0.8}{(HERA~\rom{1})}}
\newcommand{\HERAII} {\protect\scalebox{0.8}{(HERA~\rom{2})}}
\newcommand{\LowEP} {\protect\scalebox{0.8}{($300\,\GeV$)}}
\newcommand{\HLRG}  {H1 LRG \HERAII\xspace}
\newcommand{\HVFPS} {H1 VFPS \HERAII\xspace}
\newcommand{\HFPS}  {H1 FPS \HERAII\xspace}
\newcommand{\HLRGI} {H1 LRG \HERAI\xspace}
\newcommand{\ZLRG}  {ZEUS LRG \HERAI\xspace}
\newcommand{\HLRGEp}{H1 LRG \LowEP\xspace}

\newcommand{\DPDFFitA}    {H1FitA\xspace}
\newcommand{\DPDFFitB}    {H1FitB\xspace}
\newcommand{\DPDFFitJets} {H1FitJets\xspace}
\newcommand{\DPDFZSJ} {ZEUSSJ\xspace}
\newcommand{\DPDFMRW} {MRW\xspace}

\journalname{Eur. Phys. J. C}
\begin{document}

\title{Dijet production in diffractive deep-inelastic scattering in next-to-next-to-leading order QCD
}


\author{
  D.~Britzger\thanksref{eB,addrB}
  \and
  J.~Currie\thanksref{eC,addrC}
  \and
  T.~Gehrmann\thanksref{eG,addrG}
  \and
  A.~Huss\thanksref{eH,addrH}
  \and
  J.~Niehues\thanksref{eN,addrC}
  \and
  R.~\v{Z}leb\v{c}\'{i}k\thanksref{eZ,addrZ}
}

\thankstext{eB}{e-mail: britzger@physi.uni-heidelberg.de}
\thankstext{eC}{e-mail: james.currie@durham.ac.uk}
\thankstext{eG}{e-mail: thomas.gehrmann@uzh.ch}
\thankstext{eH}{e-mail: alexander.huss@cern.ch}
\thankstext{eN}{e-mail: jan.m.niehues@durham.ac.uk}
\thankstext{eZ}{e-mail: radek.zlebcik@desy.de}


\institute{Physikalisches Institut, Universit{\"a}t Heidelberg, Im Neuenheimer Feld 226, 69120 Heidelberg, Germany \label{addrB}
           \and
           Institute for Particle Physics Phenomenology, Durham University, Durham, DH1 3LE, UK \label{addrC}
           \and
           Physik-Institut, Universit\"at Z\"urich, Winterthurerstra{\ss}e 190, CH-8057 Z\"urich, Switzerland \label{addrG}
           \and
           Theoretical Physics Department, CERN, 1211 Geneva, Switzerland \label{addrH}
           \and
           DESY, Notkestra{\ss}e 85, D-22607 Hamburg, Germany \label{addrZ}
}

\date{Received: date / Accepted: date}

\maketitle

\begin{abstract}
Hard processes in diffractive deep-inelastic scattering can be described
by a factorisation into parton-level subprocesses and  
diffractive parton distributions. In this framework,
cross sections for inclusive dijet production in diffractive
deep-inelastic electron-proton scattering (DIS)  are computed to 
next-to-next-to-leading order (NNLO) QCD accuracy and compared to a
comprehensive selection of data.
Predictions for the total cross sections, 39 single-differential and four
double-differential distributions for six measurements at HERA by the H1
and ZEUS collaborations are calculated.
In the studied kinematical range, the NNLO corrections are found to be
sizeable and positive. 
The NNLO predictions typically exceed the data, while the kinematical
shape of the data is described better at NNLO than at next-to-leading
order (NLO).
A significant reduction of the scale uncertainty is achieved in
comparison to NLO predictions.
Our results use the currently available NLO diffractive parton
distributions, and the discrepancy in normalisation highlights the
need for a consistent determination of these distributions at NNLO
accuracy.

\vskip1cm
\keywords{DIS \and HERA \and Diffraction \and NNLO}
\end{abstract}

\clearpage
\section{Introduction}
\label{sec:intro}

Diffractive processes in deep-inelastic scattering, $ep \to eXY$,
where the final state systems $X$ and $Y$ are separated in rapidity,
have been studied extensively at the electron-proton collider
HERA~\cite{Newman:2013ada}.
%
The forward system $Y$ consists of the leading proton, which stays
intact after the collisions, but may also contain its low mass
dissociation.
Between the systems $X$ and $Y$ a depleted region without any
hadronic activity is observed, the so-called large rapidity gap (LRG).
This is a consequence of the vacuum quantum numbers of the diffractive
exchange which is often referred to as a pomeron ($\pom$).
Experimentally, the diffractive events can be selected either by
requiring a rapidity region in the direction of the proton beam
without any hadronic activity (LRG method) or by direct detection of
the leading proton using dedicated spectrometers.
In the second case, the system $Y$ is free of any diffractive
dissociation.

Predictions for diffractive processes in DIS can be obtained in the
framework of perturbative QCD (pQCD).
According to the factorisation theorem for diffractive DIS (DDIS)~\cite{Collins:1997sr},
if the process is sufficiently hard, the calculation can be subdivided into
two components:
the hard partonic cross sections, $\D\hat\sigma_n$, are
calculable within pQCD in powers of $\asmur$, which need to be convoluted with
soft diffractive parton distribution functions (DPDFs, $f^D_a$) that specify the
contributing parton $a$ inside the incoming hadron. 
DPDFs are universal for all diffractive deep-inelastic processes~\cite{Collins:1997sr}, 
with the hardness of the process being ensured by the virtuality $Q^2$ of the exchanged photon. 

Up to now, predictions for diffractive processes, and in particular
for diffractive dijet production, were performed only in
next-to-leading order QCD (NLO).
These predictions were able to describe the measured cross sections
satisfactorily, both in shape and normalisation 
(for a review see e.g.\ ref.~\cite{Newman:2013ada}).
However, due to their large theoretical uncertainties they did not
achieve the precision of the data and thus did not allow for more
stringent conclusions, i.e.\ about the underlying fundamental
concepts of the diffractive exchange.
Furthermore, the NLO predictions for dijet production were about two
times higher than the leading-order (LO) predictions. 
This raised the natural question concerning the size of contributions
from even higher orders for such processes at the comparably low
scales of the HERA data.
 
Here, we present the next-to-next-to-leading (NNLO) perturbative QCD
calculations for dijet production in diffractive DIS.
These calculations are performed for the first time and constitute the
first NNLO predictions for a diffractive process.
We compare our predictions with several single-, double-differential
and total cross sections from six~distinct measurements published by
the H1 or ZEUS collaboration.
A quantitative comparison of NLO and NNLO predictions with the data is
presented.
We further study the scale dependence of the NNLO predictions. 
Different DPDF parametrisations are studied and we provide additional
studies about the sensitivity of the dijet data for future DPDF determinations.

\section{NNLO predictions for dijet production in DDIS}
\label{sec:theory}
%
Relevant kinematical variables to describe fully inclusive neutral current (NC) DIS  
can be inferred from the momenta of the incoming particles and the outgoing lepton:
$$l(k)+p(P) \to l'(k') + X(p_X),$$
such that the momentum transferred to the proton is given by 
the momentum $q=k-k'$ of the virtual gauge boson $\gamma^*$.
The kinematics of each event is then completely determined by the following variables
\begin{eqnarray}
   s = (k+P)^2\,,\qquad Q^2 = -q^2\,,\qquad x = \frac{Q^2}{2q\cdot P}\,, \qquad y = \frac{q\cdot P}{k\cdot P} = \frac{Q^2}{xs}\,,                                           
\end{eqnarray}
where $y$ is referred to as the inelasticity of the scattering.
Neglecting the proton mass, the $\gamma^*p$ invariant squared mass is given by 
$W^2=sy-\Qsq$, and is thus directly proportional to $y$ in the case 
$\Qsq\ll sy$.

\begin{figure}[t]
  \centering
  \includegraphicss[width=0.4\textwidth]{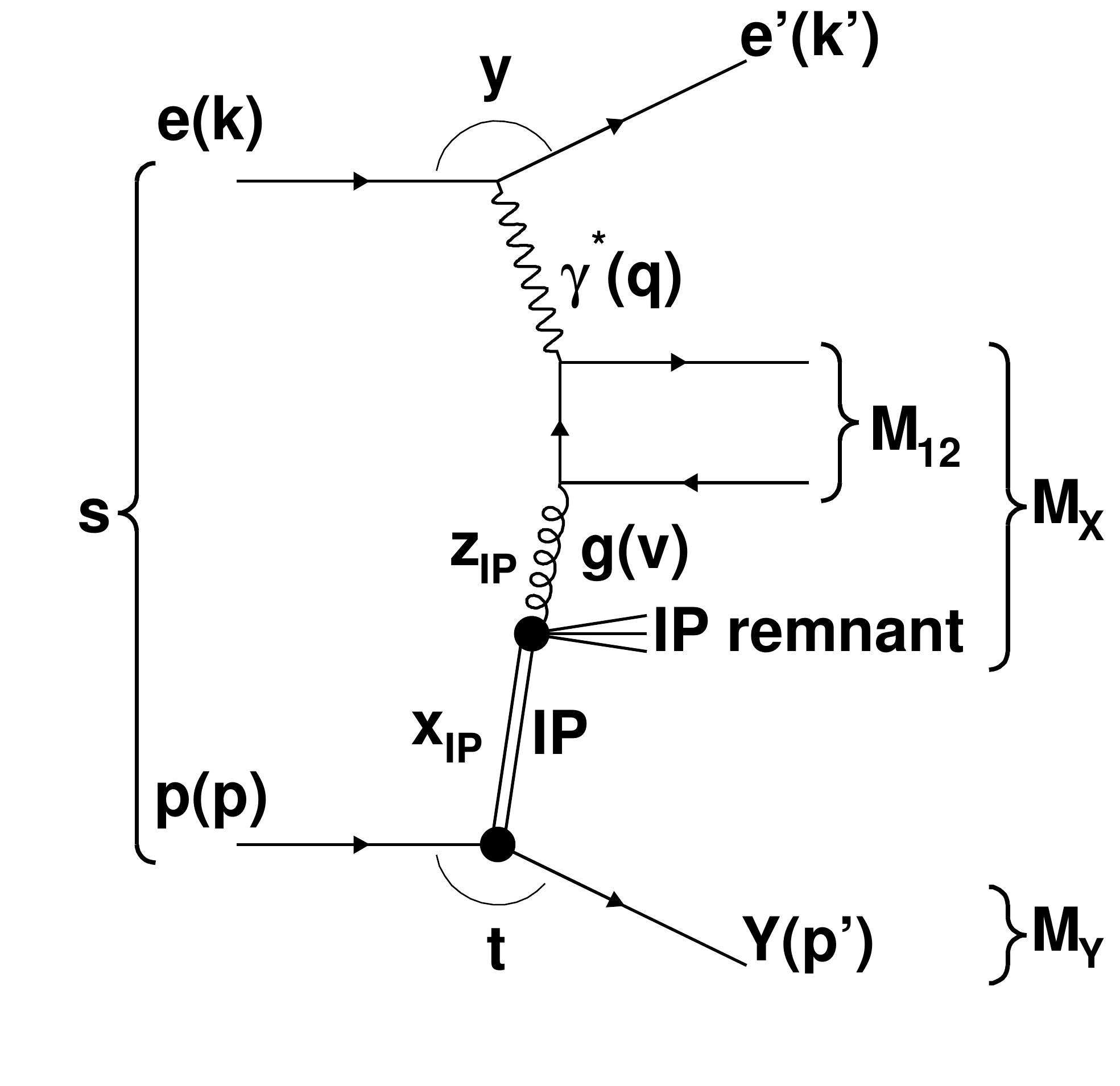}
  \caption{
    The leading order Feynman diagram for dijet production in
    diffractive DIS via boson-gluon fusion (taken from ref.~\cite{Andreev:2014yra}).
    The variables are described in the text.
  }
  \label{diagFeyn}
\end{figure}

The leading order Feynman diagram for dijet production in diffractive
DIS is displayed in figure~\ref{diagFeyn}.
In this case, a dijet system is characterised by at least two outgoing jets
 within a given pseudorapidity range (\etaj\ or \etalab)
with sufficiently high transverse momenta \ptjet\ in the
$\gamma^*p$ rest frame\footnote{Here, observables in the $\gamma^*p$
  (laboratory) frame are conventionally denoted with an asterisk
  `$^*$' (superscript `lab').}. 
At HERA, particles are commonly clustered into jets using the $k_t$ cluster
algorithm~\cite{Ellis:1993tq}.
The jet with the highest (second highest) \ptjet\  is denoted as `leading jet'
(`subleading jet') and their average transverse momentum
and invariant mass is calculated as $\ptavg=(\ptjone+\ptjtwo)/2$ and 
denoted by \mjj, respectively.

For the description of the diffractive kinematics additional
invariants have to be introduced and are in terms of the
momentum assignments from figure~\ref{diagFeyn} given by
\begin{eqnarray}
  \zpom = \frac{\mjj^2 + \Qsq}{\mxq + \Qsq}\,, \quad  
  \xpom = \frac{(p - p')\cdot q}{p\cdot q}\,, \quad
  {\rm and}\quad t = (p-p')^2\,.
  \label{zPomDef}
\end{eqnarray}
The observable \zpom\ is calculated from \mjj\ and the invariant mass of the
hadronic system $X$, $\mx$, and it characterises the parton momentum fraction 
of the diffractive exchange entering the partonic sub-process\footnote{In inclusive
  DDIS the invariant $\beta={\Qsq}/{2q\cdot (p-p')}$ has a similar
  interpretation, which can also be calculated as $\beta =
  {\xbj}/{\xpom}$, with $\xbj={\Qsq}/{2 p\cdot q}$.}. 
The denominator in the definition of \zpom\ can equivalently be written as
$\xpom ys$, i.e.\ in terms of kinematic variables related to the scattered
electron and the leading proton. 
The observable \xpom\ is interpreted as the relative energy loss of the
leading proton and is given by 
$\xpom = 1-{E_{p'}}/{E_p}$.
For measurements at HERA, \xpom\ is typically of $\mathcal{O}(0.01)$.
The variable $t$ is related to the transverse momentum of the
diffractive proton ($t \simeq -p^2_{{\rm T},p'}$) with
absolute value $\sim \! 0.1\,\text{GeV}^2$ at HERA. 
The mass of the system $Y$, which is formed by either a leading
proton or its low mass dissociative state, is denoted as \MY.


QCD predictions for sufficiently hard processes in diffractive DIS are
obtained by subdividing the calculation into two parts in accordance with
the factorisation theorem~\cite{Collins:1997sr}:
The calculation of the hard partonic scattering coefficients, $\D\hat\sigma_i$, that are
calculable within pQCD and come with the $i^{th}$ power of $\asmur$, 
and the convolution of the $\D\hat\sigma_i$ with appropriate DPDFs that capture the properties of the soft physics,
 denoted by $f^D_a$ for incoming parton of type $a$.
The full cross section up to power $n$ 
 in \asmur\ can then be written as a sum over the relevant hard coefficients
 and partonic channels,
\begin{equation}
  \sigma_n = \sum_{a=g,q,\bar{q}}\sum_{i=1}^n\sigma_{a,i}\,.
\end{equation}
In the above, the function $\sigma_{a,i}$ is calculated as a convolution 
of the DPDFs with the hard coefficients:
\begin{eqnarray}
  \sigma_{a,i} = &&\int \D t \int \D\xpom\int \D\zint\,\,\D\hat\sigma_i^{ea\rightarrow 2 {\rm jets}}(\hat{s},\mur,\muf)f^{D}_{a}(\zint,\muf, \xpom,t)\,.
  \label{eq:sigma_na}
\end{eqnarray}
Physically, the variable $\xpom$ represents the longitudinal proton momentum
fraction which contributes to the interaction or, alternatively, the
momentum fraction the proton is loosing in the diffractive exchange.
The variable \zint\ is then the fraction of the diffractive exchange
momentum which enters the hard subprocess, where it should be noted that the
variable \zint equals \zpom\ only at leading order.

The DPDFs have many properties similar to the non-diffractive PDFs,
in particular they obey the DGLAP evolution
equation~\cite{Collins:1997sr,Altarelli:1977zs,Gribov:1972ri,Dokshitzer:1977sg}, 
however, DPDFs are constrained by the presence of the leading
proton in the final state.
In parameterised DPDFs the $t$-dependence
of the cross section is integrated out and
 in the considered measurements is restricted either by $|t|<1\,\GeVsq$
or $|t|<0.6\,\GeVsq$.

In this paper, the parton-level jet-production cross sections in DDIS are calculated up to NNLO.
These calculations are identical to the NNLO calculations in the
non-diffractive case~\cite{Currie:2016ytq,Currie:2017tpe}.
The NNLO correction involves three types of
scattering  amplitudes:  the two-loop amplitudes for two-parton final
states~\cite{Garland:2001tf,Garland:2002ak,Gehrmann:2002zr,Gehrmann:2009vu}, the one-loop amplitudes for three-parton final
states~\cite{Glover:1996eh,Bern:1996ka,Campbell:1997tv,Bern:1997sc} and the tree-level amplitudes for four-parton final
states~\cite{Hagiwara:1988pp,Berends:1988yn,Falck:1989uz}.   
These contributions contain implicit infrared divergences from soft
and/or collinear real-emission corrections as well as explicit
divergences of both infrared and ultraviolet origin from the virtual
loop corrections.
When calculating predictions for an infrared-safe final state definition,
these singularities cancel when the different parton multiplicities
are combined~\cite{Sterman:1977wj}.
The  calculation  employs the antenna subtraction method~\cite{GehrmannDeRidder:2005cm,GehrmannDeRidder:2005aw,GehrmannDeRidder:2005hi,Currie:2013vh}:
For real-radiation processes, the subtraction terms are constructed
out of antenna functions, which encapsulate all color-ordered
unresolved parton emission in-between pairs of hard radiator
partons. To constitute a subtraction term, the antenna functions are
 then multiplied with reduced matrix elements of lower partonic multiplicity. 
By making the infrared pole structure explicit, the integrated
subtraction terms can be combined with the virtual corrections in
order to obtain a finite result.
Relevant tree-level and one-loop matrix elements were verified against 
Sherpa~\cite{Gleisberg:2008ta,Carli:2010cg,Cascioli:2011va} and
nlojet++~\cite{Nagy:2001xb,Nagy:2003tz}.
Our computation is performed within the parton-level event generator
\NNLOJET~\cite{Gehrmann:2018szu}, which
implements the antenna subtraction formalism and further provides a
validation framework to ensure the correctness of the results.
These tests comprise the analytic cancellation of all infrared poles
and a numerical check of the behaviour of the subtraction terms to
mimic the real-emission matrix elements in all unresolved
limits~\cite{Glover:2010im,GehrmannDeRidder:2011aa}. 
All calculations are performed using the $\overline{\rm MS}$
renormalisation scheme and for five massless quark flavors.
The strong coupling constant is set to $\asmz=0.118$~\cite{Patrignani:2016xqp}.

The calculation of NNLO partonic cross sections~\cite{Currie:2016ytq,Currie:2017tpe} has recently been 
applied successfully 
to describe inclusive jet and dijet cross
section data in non-diffractive DIS~\cite{Currie:2016ytq,Andreev:2016tgi,Currie:2017tpe,Andreev:2017vxu}.
Here, however, the hard coefficients are now convoluted with DPDFs for the first time and
we present the first calculation of a diffractive jet production process to NNLO in \asmur.
For this reason our predictions are limited by the available DPDFs which have
 only been determined up to NLO so far.
While previous calculations of NLO diffractive dijet cross sections
commonly used the computationally very expensive slicing method~\cite{Aktas:2007bv},
 here an improved convolution formalism is used.
Our calculation thereby employs the fastNLO
formalism~\cite{Andreev:2014yra,Britzger:2012bs,Britzger:2013kia} which has the
advantage that the matrix elements have to be calculated only once and
can then be used repeatedly for integrations of the DPDFs.
The formalism will be briefly explained in the following.

The matrix elements
$\D\hat\sigma_n^{ea\rightarrow 2 {\rm jets}}$ have their \xpom\ and
\zint\ dependence given through $\hat{s}= xs$, where $s$ is the
centre-of-mass energy squared of the $ep$ collision and the momentum 
fraction $x$ is given by 
\begin{equation}
  x = \xpom\zint\,.
  \label{eq:x}
\end{equation}
In the fastNLO approach for non-diffractive DIS the $x$-dependence of
the matrix elements is frozen on a grid,
\begin{equation}
  \int \D x\, \D\hat\sigma_{a,n}(x)\, f(x) \simeq \sum_i \tilde{\sigma}_i^{(a,n)} f(x_i)
  \label{eq:fnloddis}
\end{equation}
where the nodes lie at set values of $x_i$. With an increasing number
of nodes the approximation improves until both expressions in eq.~\eqref{eq:fnloddis}
become numerically identical.
The coefficients $\tilde\sigma_i$ are calculated from contributing matrix
elements for a given measurement function, which expresses the given
observable, phase space and jet definition.
While this calculation is computationally very expensive, it has to be
performed only once, since
these coefficients are 
independent of PDF values (DPDFs) and scales. 

Using eq.~\eqref{eq:fnloddis} the partonic cross section in DDIS
eq.~\eqref{eq:sigma_na} is then calculated as
\begin{equation}
  \sigma_{a,n} =
  \int \D t\int \tfrac{\D\xpom}{\xpom} \sum_i^{x_i<\xpom} \tilde\sigma_i^{(a,n)} {f}^D_a(\xpom,\zint=x_i/\xpom,t)\,.
  \label{eq:sigmafnlo}
\end{equation}
By interpreting the factor $1/\xpom$ as the flux factor of the
diffractive exchange, then according to a center-of-mass reweighting of the incoming hadron, the
calculation can be made equivalent to the slicing method.
Our calculations have been validated in NLO accuracy against
calculations using nlojet++~\cite{Nagy:2001xb,Nagy:2003tz} with the slicing method. 

The fastNLO based approach has advantages of a higher numerical accuracy of the
\xpom\ integration, and,
more importantly still, a significantly higher numerical accuracy is achieved in the
calculation of the hard matrix elements for a given amount of computing time.
This is of great importance for the calculation of the
double-real and real-virtual NNLO amplitudes, which are calculated
here using several 100,000~hours of CPU time using state-of-the-art CPUs.
The numerical accuracy of the fastNLO interpolation technique is
typically smaller than the numerical precision of the tabulated
DPDFs, and thus can be neglected.

In order to avoid regions of the phase space where the predictions
exhibit an enhanced infrared
sensitivity~\cite{Potter:1999gg,Duprel:1999wz}, the phase space
definitions of all analyses have asymmetric cuts on the transverse 
momenta of the two leading jets.
It was tested that the difference of $\sim\! 1\, \text{GeV}$ between
the cuts on the leading and sub-leading jet is sufficient to remove
this region. 

For the nominal calculations the renormalisation (\mur) and
factorisation scale (\muf) are set to 
\begin{equation}
  \mur^2=\muf^2=\Qsq+\ptavg^2\,,
\end{equation}
while also different choices are studied.
The `scale' uncertainty of the prediction is obtained by varying \mur\ and
\muf\ by the conventional factors of 0.5 and 2.

Diffractive parton distributions are determined by interpreting 
data for different final states in DDIS in a parton model framework~\cite{Ingelman:1984ns}. 
Already the first inclusive DDIS data from HERA~\cite{Ahmed:1995ns} 
indicated the presence of a very large gluon content in the diffractive exchange~\cite{Gehrmann:1995by}.
The knowledge of the DPDFs is at a lower precision than that of non-diffractive
PDFs. 
This is due to the uncertainties of the DDIS measurements, but also because 
available data sets are not always compatible~\cite{Aaron:2012ad}.
In addition, different assumptions imposed for their determination result in
substantial differences of individual DPDFs.
Therefore, different DPDF sets may result in sizeable differences for
certain processes and kinematic regions.
Currently, all DPDFs available have been obtained using data together
with corresponding NLO QCD predictions only.
Given the typical scales of the HERA measurements, higher order QCD
effects are sizable and NNLO DPDFs are expected to differ significantly
from their NLO variants.
Nonetheless, due to the absence of NNLO DPDFs we have to use NLO DPDFs
and the following sets are studied:
\begin{compactitem}
  \item \DPDFFitB~\cite{Aktas:2006hy} is the most widely used DPDF.
    It was determined from an NLO DGLAP 
    QCD fit to reduced inclusive DDIS cross sections.
    The diffractive data was selected using the LRG method and, therefore,
    the DPDF includes proton dissociation into a low-mass hadronic state
    ($\MY<1.6\,\GeV$). 
    The phase space of the selected data was restricted to $\beta < 0.8$
    and $\Qsq>8.5\,\GeVsq$. 
    The gluon DPDF at the starting scale of the evolution,
    $\mu_0^2 = 1.75\,\GeVsq$,
    was assumed to be a constant, i.e.\ independent of the value of \zint.   
  \item \DPDFFitA~\cite{Aktas:2006hy} is a variant of the \DPDFFitB DPDF,
    which uses a more flexible parametrisation of the gluon
    distribution at the starting scale of the evolution.
    In comparison to the \DPDFFitB DPDF, a significantly larger gluon DPDF is
    found although both, the \DPDFFitA and the \DPDFFitB DPDF, describe the 
    shape of the data equally well, as inclusive
    DDIS cross sections are only weakly sensitive to the gluon DPDF.
    A detailed analysis of dijet data suggests~\cite{Aktas:2007bv}
    that the gluon component in the \DPDFFitA DPDF is overestimated.
  \item \DPDFFitJets~\cite{Aktas:2007bv} is the first DPDF fitted based on the combination of inclusive and dijet data, using the same inclusive data sample as for \DPDFFitB and \DPDFFitA .
    The inclusion of dijet data, which is more sensitive to the gluon content, led to a slightly smaller gluon distribution compared to the \DPDFFitB DPDF.
  \item \DPDFZSJ~\cite{Chekanov:2009aa} is determined by a combined fit of inclusive and dijet data by the ZEUS collaboration.
    Compared to H1 fits, the proton dissociation has been subtracted
    using Monte Carlo (MC) estimates such that this DPDF is defined
    for elastic scattering ($\MY=m_P$). 
  \item The \DPDFMRW DPDF~\cite{Martin:2006td} is based on the same data as
    the \DPDFFitB DPDF. In contrast, however, Regge factorisation is only assumed at the starting
    scale and the evolution is performed using inhomogeneous evolution
    equations accounting for pomeron-to-parton splittings.
\end{compactitem}
The DPDF uncertainty in our calculations is obtained from the error
sets provided together with the \DPDFFitB DPDF.
The very recent GKG18 DPDF~\cite{Goharipour:2018yov}, which is also in
NLO, is not considered in this analysis.

Similarly as in the definitions for DPDF fits, also the various measurements impose different
definitions of \MY.
The LRG measurements by H1 are defined for $\MY<1.6\,\GeV$, whereas
ZEUS extrapolated its LRG measurement to $\MY=m_P$. 
Two of the H1 measurements are based on proton spectrometers (FPS, VFPS), and thus
these data do not contain any proton dissociation ($\MY=m_P$). 

In order to provide predictions for all of the measured cross section data
with any of the available DPDF sets, correction factors for proton dissociation have to be applied whereever applicable.
The latest value of the proton dissociation fraction for the phase
space imposed by H1 was estimated to be~\cite{Aaron:2010aa} 
\begin{equation}
\frac{\sigma ( \MY < 1.6\,\text{GeV})}{\sigma ( \MY = m_P)} = 1.20 \pm 0.11 (\mathrm{exp.})\,.
\end{equation}
This value was obtained as a combination of the previously measured
value of $1.23 \pm 0.16$~\cite{Aktas:2006hx} and a newly measured
value of $1.18 \pm 0.12$~. 
It is consistent with the prediction of $1.15$~ obtained with the DIFFVM generator~\cite{List:1998jz}.

In order to compare the data with fixed-order predictions, correction
factors accounting for hadronisation effects are applied.
These are estimated using MC simulations and corresponding correction factors are
provided together with the respective data as discussed in the next section.

\section{Data sets and observables}
\label{sec:results}
%
The NNLO cross sections are computed for six
measurements taken at HERA by the H1 or ZEUS collaborations. 
We will refer to them as
\begin{compactitem}
\item \HFPS ~\cite{Aaron:2011mp},
\item \HVFPS~\cite{Andreev:2015cwa},
\item \HLRG ~\cite{Andreev:2014yra},
\item \HLRGI~\cite{Aktas:2007bv},
\item\HLRGEp~\cite{Aktas:2007hn}, and
\item\ZLRG~\cite{Chekanov:2007aa}.
\end{compactitem}
Five of those are performed at a centre of mass energy of
$\sqrt{s}=319\,\GeV$, and one at $\sqrt{s}=300\,\GeV$~\cite{Aktas:2007hn}, 
depending on the proton beam energy of 920\,\GeV\ or 820\,\GeV, respectively, while
the electron or positron beam energy was always equal to 27.6\,\GeV.
In two cases the leading proton is identified by the Forward Proton
Spectrometer (FPS)~\cite{Aaron:2011mp} or Very Forward Proton
Spectrometer (VFPS)~\cite{Andreev:2015cwa}, otherwise the diffractive
events are selected using the LRG method. 
Jets were identified using the $k_T$ jet
algorithm in the $\gamma^*p$ frame with cone parameter
$R=1$, and at least two jets are required in each event.
The phase space definitions of the measurements are summarised in
table~\ref{tab:datasetsInfo}.
The hadronisation corrections are provided together with the
data~\cite{Andreev:2014yra,Aktas:2007bv,Aaron:2011mp,Andreev:2015cwa,Aktas:2007hn}, 
or in case ref.~\cite{Chekanov:2007aa},  are displayed in ref.~\cite{Bonato:2008zz}.
Dijet cross sections are studied differentially in several kinematic
variables, which also constrain the phase space of the measurements, and
their meanings are described in figure~\ref{diagFeyn}.
\begin{table*}[tbhp]
  \scriptsize
    \caption{
      Summary of the dijet data sets. 
      The first column represents the data set label and the second
      shows the integrated luminosity and the number of events of the
      given data set. 
      The other columns summarise the definition of the phase space of
      the given data.
      In cases, where the DIS phase space is defined in terms of $W$,
      the corresponding range in $y=W^2/s$ is shown. 
      All measurements have in common a requirement of $n_{\rm jets}
      \geq 2$, which is applied after identifying the two leading
      jets.}
    \label{tab:datasetsInfo}
  \begin{center}
    \begin{tabular}{ccccc}
            \hline
            {\bf Data Set}  & {\boldmath{$\mathcal{L}$}}
            & {\bf DIS} & {\bf Dijet} & {\bf Diffractive}   \\     %
            & [{\invpb}]
            & {\bf range} & {\bf range} & {\bf range} \\
      \hline
      \HFPS \cite{Aaron:2011mp} &  156.6 &  $4<\Qsq<110\,\GeVsq$ & $\ptjone>5\,\GeV$ & $\xpom<0.1$ \\
            &(581ev)& $0.05 < y < 0.7$ & $\ptjtwo>4.0\,\GeV$& $|t|<1\,\GeVsq$ \\
            & & & $-1<\etalab<2.5$& $\MY= m_P $ \\
      \hline
      \HVFPS \cite{Andreev:2015cwa} &  50 &  $4<\Qsq<80\,\GeVsq$ & $\ptjone>5.5\,\GeV$ & $0.010<\xpom<0.024$ \\
            &(550ev)& $0.2 < y < 0.7$ & $\ptjtwo>4.0\,\GeV$& $|t|<0.6\,\GeVsq$ \\
            & & & $-1<\etalab<2.5$ & $\MY= m_P $ \\
      \hline
      \HLRG \cite{Andreev:2014yra} &  290 &  $4<\Qsq<100\,\GeVsq$ & $\ptjone>5.5\,\GeV$ & $\xpom<0.03$ \\
            &($\sim$15000ev)& $0.1 < y < 0.7$ & $\ptjtwo>4.0\,\GeV$& $|t|<1\,\GeVsq$ \\
            & & & $-1<\etalab<2$ & $\MY<1.6 \, \GeV$ \\
      \hline
      \HLRGI \cite{Aktas:2007bv} & 51.5 &  $4<\Qsq<80\,\GeVsq$ & $\ptjone>5.5\,\GeV$ & $\xpom<0.03$ \\
            &(2723ev)& $0.1 < y < 0.7$ & $\ptjtwo>4.0\,\GeV$& $|t|<1\,\GeVsq$ \\
            & & & $-3<\eta^{*\mathrm{jet}}<0$ & $\MY<1.6 \, \GeV$ \\
      \hline
      \HLRGEp \cite{Aktas:2007hn} &  18 &  $4<\Qsq<80\,\GeVsq$ & $\ptjone>5\,\GeV$ & $\xpom<0.03$ \\
             &(322ev)& $165 < W < 242\,\GeV$ & $\ptjtwo>4.0\,\GeV$& $|t|<1\,\GeVsq$ \\
             & & $(0.30 < y < 0.65)$& $-1<\etalab<2$& $\MY<1.6 \, \GeV$ \\
             & &  & $-3<\eta^{*\mathrm{jet}}<0$&  \\
      \hline
      \ZLRG \cite{Chekanov:2007aa} &  61 &  $5<\Qsq<100\,\GeVsq$ & $\ptjone>5\,\GeV$ & $\xpom<0.03$ \\
            &(5539ev)& $100 < W < 250\,\GeV$ & $\ptjtwo>4.0\,\GeV$& $|t|<1\,\GeVsq$ \\
            & &  $(0.10 < y < 0.62)$  & $-3.5<\eta^{*\mathrm{jet}}<0$ & $\MY = m_P$ \\
      \hline
    \end{tabular}
    \end{center}
\end{table*}

Measurements were performed as functions of:
\begin{compactitem}
  \item The DIS kinematic variables:
    \Qsq, $y$ and $W$;
  \item The jet transverse momentum observables: \ptjone, \ptjtwo,
    \ptavg and \ptjet. Here \ptjet refers to the $p_\mathrm{T}$ of the
    leading and subleading jet;
  \item The jet pseudorapidity observables: \etaavg, \etaj, \deleta,
    and \deletastar. Here \etaavg denotes the average
    pseudorapidity \etaj of the two leading jets and \deleta\ and
    \deletastar\ denote their separation in pseudorapidity;
  \item Observables of the diffractive final state:  
    \xpom, \zpom and \mx;
  \item Double-differential measurements as functions of \zpom
     or \ptjone for \Qsq  intervals, and
  as a function of \zpom for \ptjone  intervals.
\end{compactitem}

In the fastNLO approach, the  $\tilde{\sigma}_i^{(a,n)}$ coefficients
are calculated prior to the convolution with the DPDFs.
In this first step, however, only observables that are accessible from information 
on the final state kinematics of the hard matrix element can be evaluated directly.
An example for such variables are the DIS kinematic variables or jet momenta.
In contrast, the kinematics of the hard matrix elements do not depend
explicitly on the outgoing  proton momentum.
Observables depending on the diffractive final state have therefore to be derived
in additional steps when the \xpom and $|t|$ integration is performed
(c.f.\ eq.~\eqref{eq:sigmafnlo}). 
In such cases (for instance for the \xpom and $|t|$ observables),
differential predictions are obtained from 
$\tilde{\sigma}_i^{(a,n)}$ coefficients representing the total hard cross section.
Similarly, predictions as a function of \zpom are calculated using the relation
$\zpom=\xi/\xpom$ and are obtained from $\tilde{\sigma}_i^{(a,n)}$ coefficients 
for a highly resolved distribution in $\xi$, which denotes the proton momentum fraction carried  by the incoming parton at leading order and is calculated as $\xi=\xbj(1+\mjj^2/\Qsq)$~\cite{Currie:2017tpe}.
Predictions as a function of \mx are obtained using the $\tilde{\sigma}_i^{(a,n)}$ coefficients
for a highly resolved distributions in $y$ and $Q^2$, in combination with $\mx = \sqrt{ys\xpom -\Qsq}$.

\section{Results}
\subsection{Total dijet production cross section}
\label{sec:total}
\begin{figure*}[h]
\centering
\includegraphicss[width=0.6\linewidth]{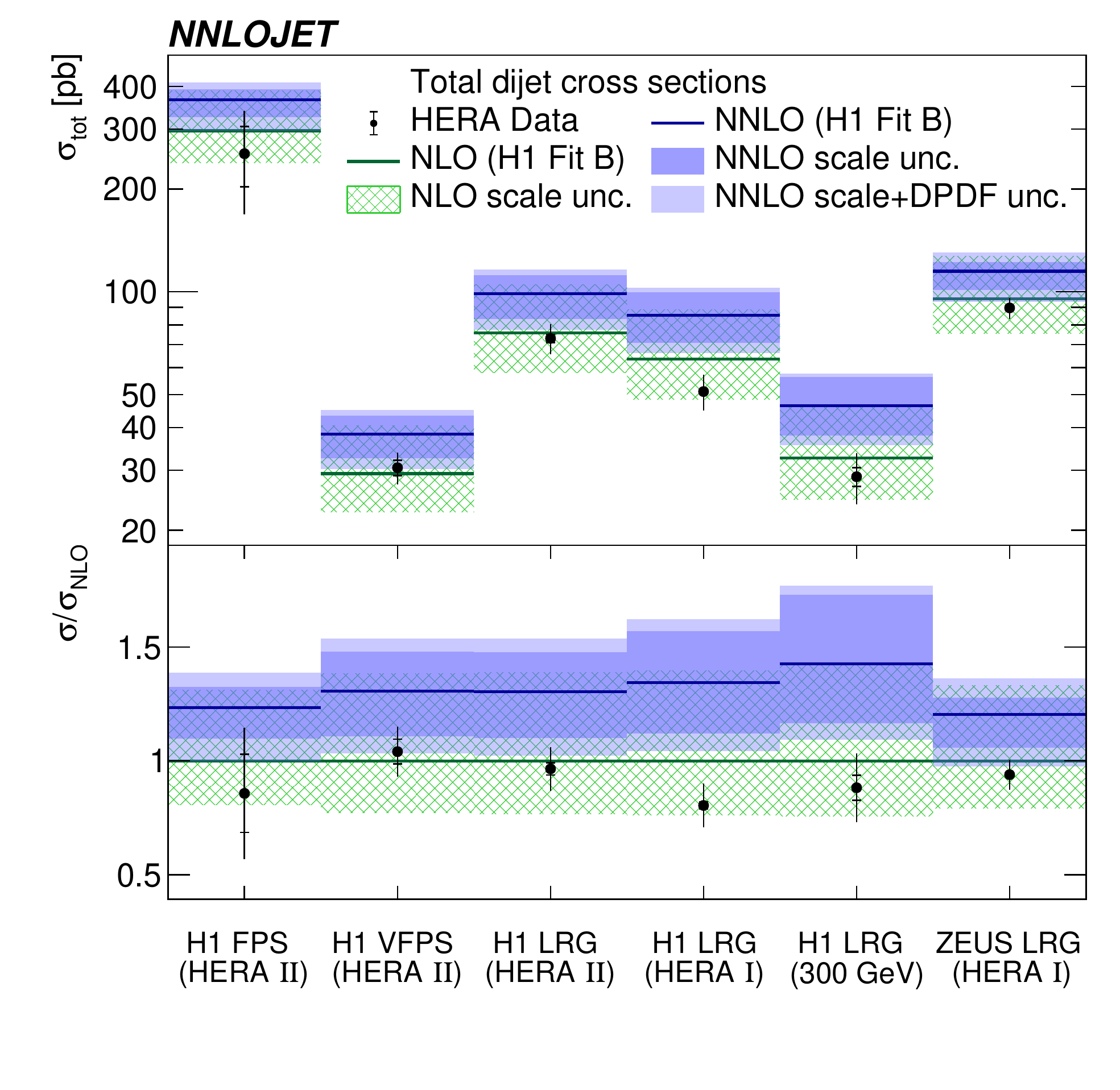}
\caption{The comparison of the QCD predictions at NLO and NNLO for the
  total dijet cross sections with the measurements.
  The inner data error bars represent statistical
  uncertainties and other error bars are statistical and systematic
  errors added in quadrature. The theoretical predictions
  using \DPDFFitB are displayed together with their scale uncertainties (NLO
  and NNLO) and with scale and DPDF uncertainties added in quadrature
  (only NNLO).
  The lower panel displays the ratio to the NLO predictions.
} 
\label{fig:tot:NLOvsNNLO}
\end{figure*}

The NNLO predictions for the total dijet cross sections of the six different experimental
measurements are presented in table~\ref{tab:NNLOtot} and are
graphically displayed in 
figure~\ref{fig:tot:NLOvsNNLO}. In both, results for the corresponding measured 
cross sections as well as for the NLO predictions are also included.
\begin{table*}[tbh]
  \renewcommand{\arraystretch}{1.6} 
  \caption{
    Comparison of the measured and predicted total dijet cross sections for
    the six measurements.
    Listed are the data cross section, $\sigma^{\rm Data}$, the NLO
    and the NNLO predictions, $\sigma^{\rm NLO}$ and
    $\sigma^{\rm NNLO}$, respectively.
    For $\sigma^{\rm Data}$ the uncertainties denote the
    statistical  and the systematic  uncertainty.   
    In case of \HLRGEp, the total cross section is
    calculated by us from the single-differential distributions.
    The uncertainty of the NLO or NNLO predictions denote the scale
    uncertainty obtained from a simultaneous variation of \mur\ and
    \muf\ by factors of 0.5 and 2.
    The last two columns show the DPDF uncertainty obtained from 
    \DPDFFitB for the NLO or NNLO predictions.
    In terms of a relative uncertainty, the DPDF uncertainty is almost
    identical for NLO and NNLO predictions.
    }
    \label{tab:NNLOtot}
  \scriptsize
  \begin{center}
    \begin{tabular}{lccccc}
      \hline
            {\bf Data set} & $\sigma^{\rm Data}$ & $\sigma^{\rm NLO}$ & $\sigma^{\rm NNLO}$ & $\Delta_{\rm DPDF}^{\rm NLO}$ & $\Delta_{\rm DPDF}^{\rm NNLO}$ \\
            {\bf }  & [pb] & [pb] & [pb] & [pb] & [pb] \\
            \hline
            \HFPS         & $254\pm20\pm27$  
            & $296^{+92}_{-57}$  & $366^{+27}_{-41}$ &
            $^{+29}_{-46}$ &$^{+36}_{-57}$  \\
            \HVFPS        & $30.5\pm1.6\pm2.8$  
            & $29.3^{+11.2}_{-6.7}$  & $38.3^{+5.1}_{-5.8}$  
            & $^{+3.2}_{-4.2}$ &$^{+4.4}_{-5.6}$\\
            \HLRG         & $73\pm2\pm7$  
            & $75.7^{+29.4}_{-17.7}$  & $98.6^{+13.2}_{-15.4}$  
            &$^{+8.5}_{-10.9}$ &$^{+11.7}_{-14.7}$ \\
            \HLRGI         & $51\pm1^{+7}_{-5}$  
                        & $63.4^{+25.2}_{-15.1}$  & $85.3^{+14.3}_{-14.3}$  
            &$^{+7.1}_{-9.2}$ &$^{+10.1}_{-12.7}$ \\
            \HLRGEp & $28.7\pm 1.8 \pm 3.0$
            & $32.5^{+13.7}_{-7.9}$  & $46.4^{+9.9}_{-8.5}$  
            &$^{+3.5}_{-4.6}$ &$^{+5.3}_{-6.7}$ \\
            \ZLRG       & $89.7\pm1.2^{+6.0}_{-6.4}$ 
            & $95.5^{+31.5}_{-20.0}$  & $114.9^{+7.1}_{-13.8}$  
             &$^{+10.5}_{-13.4}$ &${}^{+13.5}_{-16.7}$ \\
      \hline
    \end{tabular}
    \end{center}
\end{table*}
The NNLO predictions compared to the NLO predictions are higher by about 20 to 40\,\%.
Since the kinematic ranges of different measurements are rather similar
(table~\ref{tab:datasetsInfo}), also the  
NNLO corrections are of similar size for the individual
measurements.
As found previously~\cite{Andreev:2014yra,Aktas:2007bv,Aaron:2011mp,Andreev:2015cwa,Aktas:2007hn},
the NLO predictions provide a good description for all of the
data.
In contrast, the NNLO predictions typically overshoot the data.
This tension between NNLO and data may be attributed to inappropriate
DPDFs, where we use the \DPDFFitB DPDF set, which has been determined using
NLO predictions.
In particular, the gluon component in this DPDF appears to be
too high for the usage with NNLO QCD coefficients, as this DPDF has been
determined from inclusive DDIS cross section data using the
respective NLO predictions only.


When compared to our common predictions, all measurements appear to be
consistent with each other, although they use different techniques for the
identification of the diffractive final states.

\subsection{NNLO scale uncertainty and scale choice}
The scale uncertainties, which are obtained by a simultaneous variation of
\mur\ and \muf\ by factors of $0.5$ and $2$,
are found to be reduced significantly for NNLO predictions in
comparison to NLO predictions
(see also table~\ref{tab:NNLOtot} and figure~\ref{fig:tot:NLOvsNNLO}).
The typical size of the scale uncertainty of the total dijet cross sections at NNLO is about 15\,\%,
whereas it is about  35\,\% in NLO.
In case of the \HLRG\ total cross section for instance, the upward
(downward) scale uncertainty is reduced from 39\,\% (23\,\%) at NLO to 13\,\%
(16\,\%) at NNLO.
This makes these uncertainties competitive with the data uncertainty ($\sim\! 10\%$).
For all total cross section measurements, however, the differences
between data and NNLO predictions are larger than respective
theoretical scale uncertainties.

\begin{figure}[tbh]
\centering
\begin{minipage}{.49\textwidth}
\includegraphicss[width=1.0\textwidth]{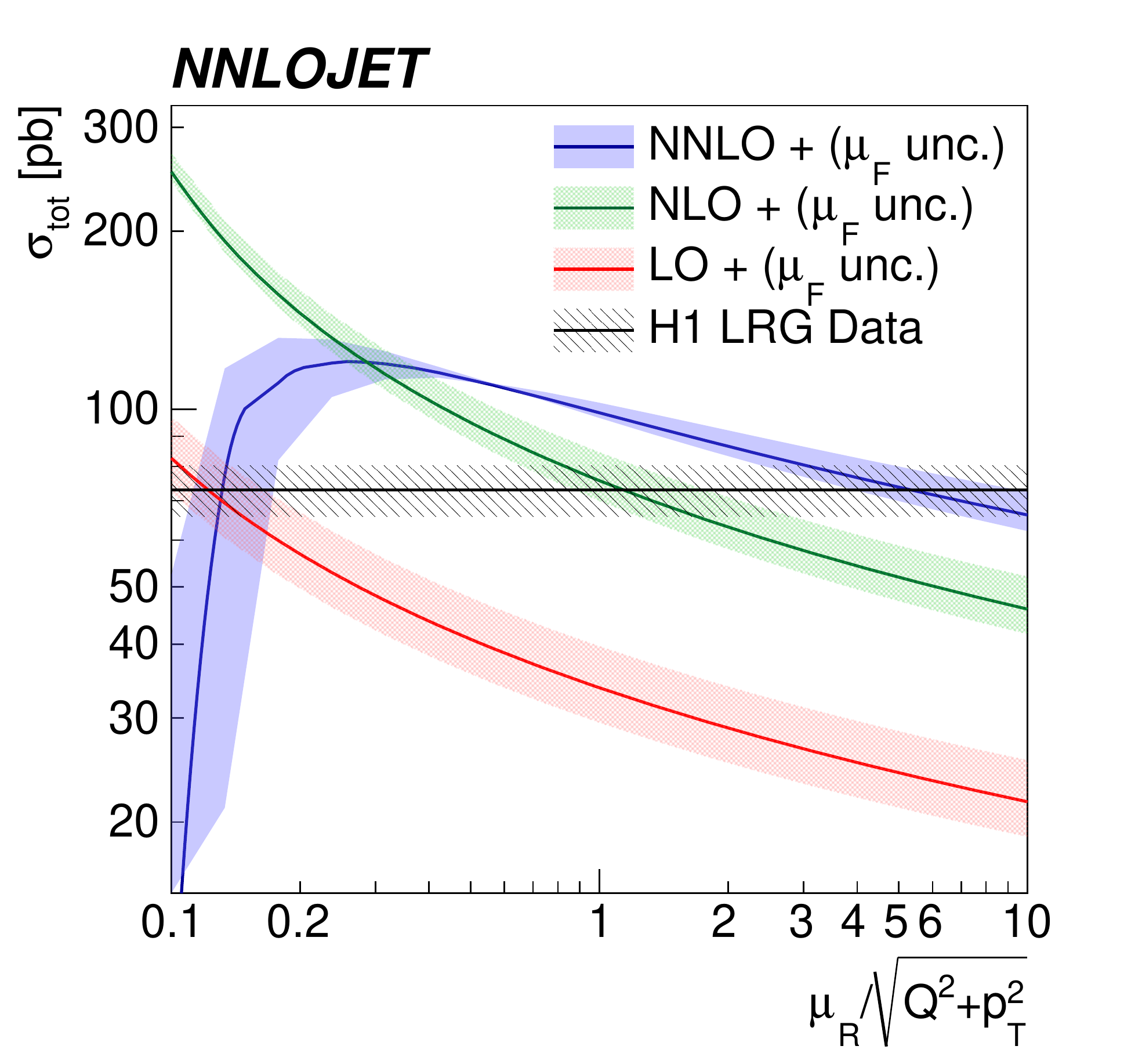}
\end{minipage}
\begin{minipage}{.49\textwidth}
\includegraphicss[width=1.0\textwidth]{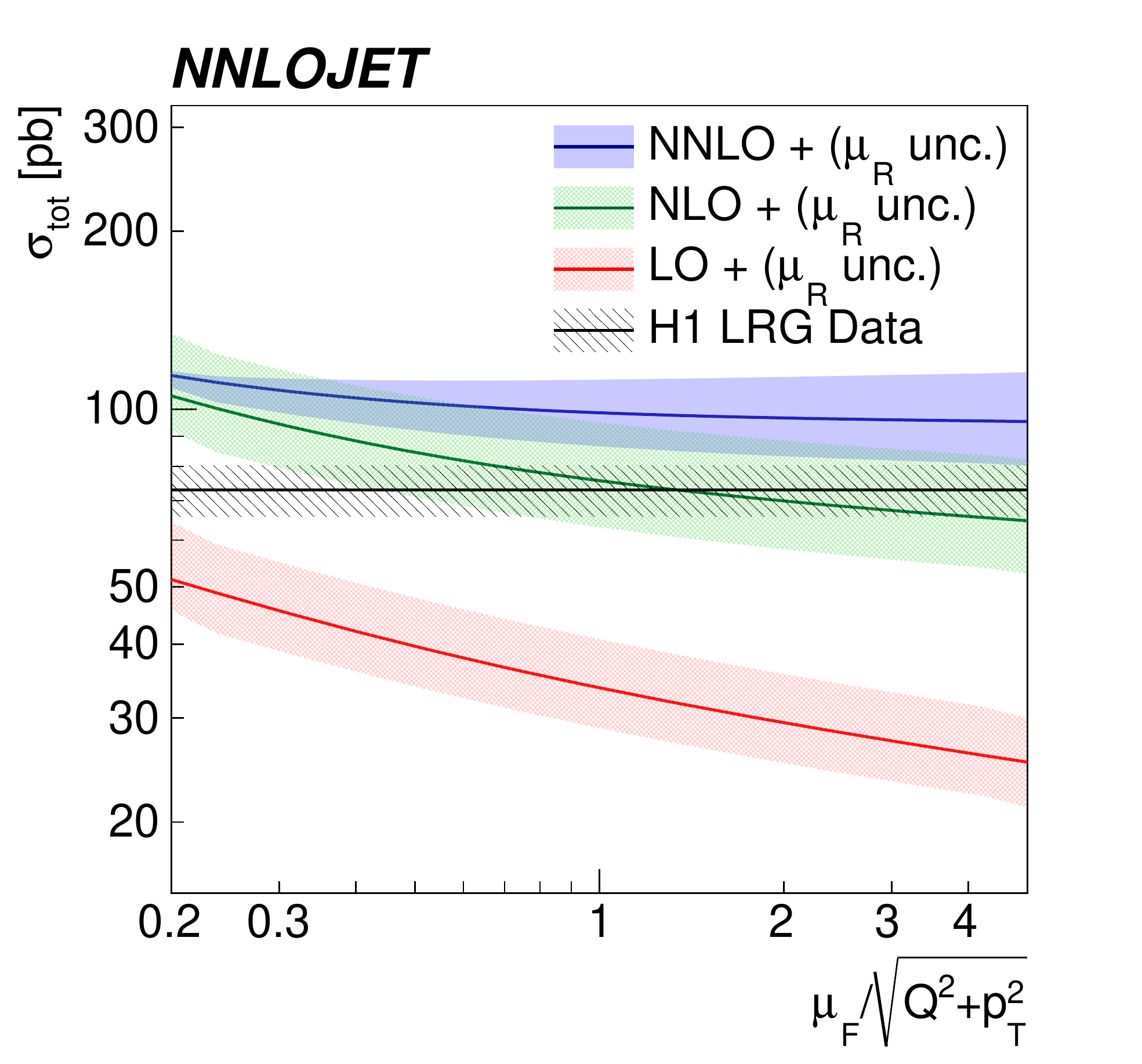}
\end{minipage}
\caption{ The dependence of the total dijet cross section of the \HLRG
  analysis on the renormalisation
  (left) and factorisation (right) scale.  
  The left (right) panel displays a variation of \mur\ (\muf) by
  factors between $0.1$ and $10$ and the 
  effect of the variation of \muf\ (\mur) with factors of $0.5$ and
  $2$ is displayed by the shaded areas.
  The calculated cross sections are shown at LO, NLO and NNLO accuracy. 
  The measured data cross section with its total
  uncertainty is displayed as a black line and hatched area.
} 
\label{fig:ScaleDependence}
\end{figure}

A detailed investigation of the scale dependence of the LO, NLO
and NNLO predictions is displayed in figure~\ref{fig:ScaleDependence}
for the \HLRG phase space.
While the NLO scale dependence is of similar size as for LO predictions,
 the scale dependence of the NNLO predictions is significantly reduced.
The \mur\ dependence is significantly larger than the
\muf\ dependence, which is also found for non-diffractive jet
production~\cite{Andreev:2017vxu}.
The K-factor of the NNLO correction (defined as $\sigma_{\rm NNLO}/\sigma_{\rm NLO}$) is found to
be significantly smaller than the K-factor of the NLO corrections 
($\sigma_{\rm NLO}/\sigma_{\rm LO}$), thus indicating convergence of the perturbative series.
In comparison to data, the NNLO predictions exceed the \HLRG data for a wide range of scale factors.

\begin{figure}[tbh]
  \centering
  \begin{minipage}{.49\textwidth}
    \includegraphicss[width=1.0\textwidth]{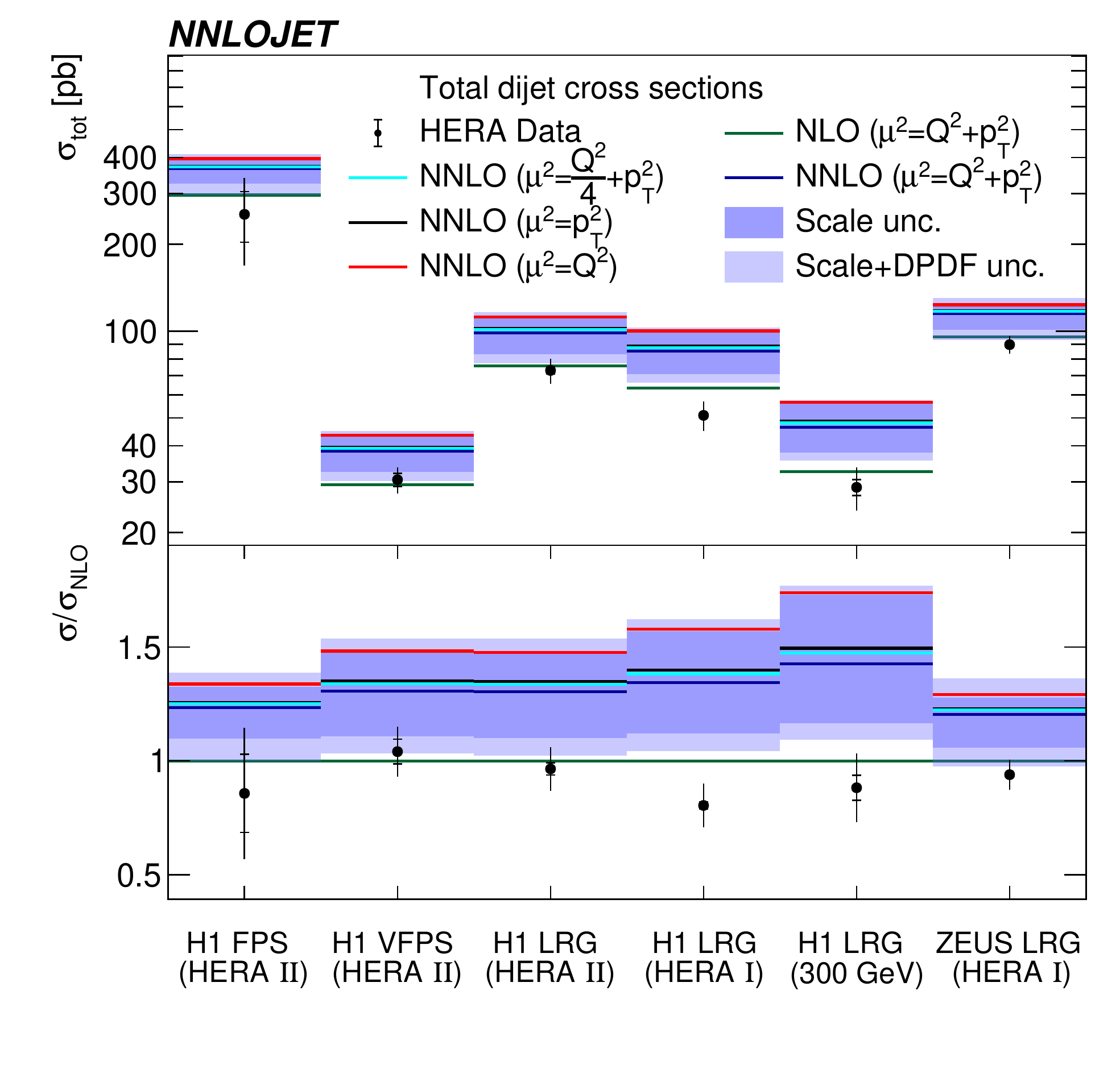}
  \end{minipage}
  \begin{minipage}{.49\textwidth}
    \includegraphicss[width=1.0\textwidth]{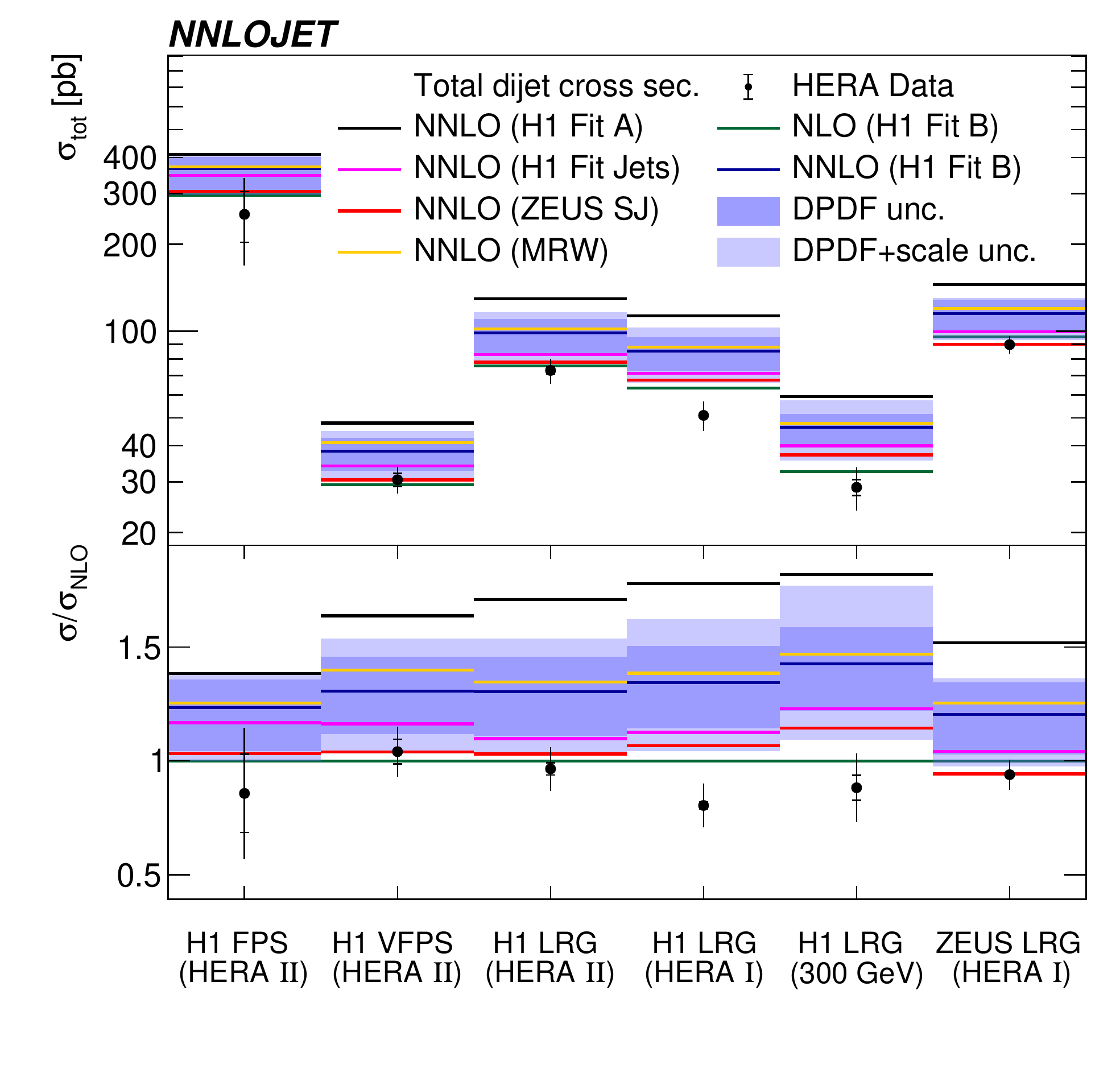}
  \end{minipage}
  \caption{
    The comparison of the NNLO predictions for the
    total dijet cross sections with the measurements and NLO predictions.
    The dark shaded bands display the scale (left) and DPDF
    uncertainties (right), and the light shaded bands display these
    uncertainties added in quadrature.
    The left panel displays NNLO predictions for different scale
    definitions.
    The right panel displays NNLO predictions for different DPDF
    choices.
    The lower panels display the ratio to NLO predictions.
  }
  \label{fig:TotXsec2}
\end{figure}

The NNLO calculations are repeated for alternative
choices for $\mur^2$ and $\muf^2$ using $\tfrac{\Qsq}{4}+\ptavg^2$, 
$\ptavg^2$ and $\Qsq$,
and results are displayed in figure~\ref{fig:TotXsec2} (left).
Numerical values for the phase space of the \HLRG\ analysis are listed in
table~\ref{tab:Scales}.
The cross sections obtained with scale choices involving
$\ptavg^2$ in their definitions differ only moderately among each other.
In contrast, a scale choice of $\mu^2=\Qsq$ changes the predictions significantly 
compared to the aforementioned scale choices. In this case, the 
differences are of similar size to the scale uncertainties. 
This can be traced back to kinematic regions where \Qsq\ is small compared to $\ptavg^2$,
and a choice of \Qsq\ can be considered as inappropriate.
\begin{table*}[tbh]
    \caption{
      NNLO predictions for \HLRG using different
      choices for $\mur^2$ and $\muf^2$.
      The uncertainties denote the scale uncertainty from
      simultaneously varying \mur\ and \muf\ by factors of 0.5 or 2.
    }
    \label{tab:Scales}
  \scriptsize
  \begin{center}
    \begin{tabular}{lcccccc}
            \hline
            {\bf Data set} & $\sigma^{\rm Data}$ 
            & $\Qsq+\ptavg^2$
            & $\Qsq$
            & $\ptavg^2$
            & $\tfrac{\Qsq}{4}+\ptavg^2$
            & $\sqrt{Q^4+\ptavg^4}$ 
            \\
            & [pb] & [pb] & [pb] & [pb] & [pb]& [pb] \\
            \hline
            \HLRG & 
            $73\pm7_{\rm exp}$  
            & $98.6^{+13.2}_{-15.4}$  
            & $111.7^{-43.4}_{-11.5}$  
            & $102.1^{+8.4}_{-15.2}$  
            & $101.1^{+10.6}_{-15.4}$  
            & $101.0^{+11.2}_{-15.5}$   \\
      \hline
    \end{tabular}
    \end{center}
\end{table*}

\subsection{DPDF choice and uncertainties}
In figure~\ref{fig:TotXsec2} (right), we study the dependence of the total 
cross sections on the choice of DPDFs, using
\DPDFFitA~\cite{Aktas:2006hy}, 
\DPDFFitB~\cite{Aktas:2006hy}, 
\DPDFFitJets~\cite{Aktas:2007bv},
\DPDFMRW~\cite{Martin:2006td}
and
\DPDFZSJ~\cite{Chekanov:2009aa} DPDFs. 
Numerical values for the \HLRG phase space are provided in table~\ref{tab:DPDFs}.
The NNLO predictions overshoot the data for any choice of DPDFs.
However, it is observed that DPDFs that also consider dijet data in their
determination~\cite{Aktas:2007bv,Chekanov:2009aa} (using dijet
NLO predictions) give smaller
predictions than DPDFs that depend on inclusive DDIS data only~\cite{Aktas:2006hy}.  
The differences between the predictions are mostly covered by the DPDF
uncertainties of \DPDFFitB. 
The DPDF \DPDFFitA~\cite{Aktas:2006hy} predicts a much larger
cross section and thus appears to overestimate the gluon
component significantly. 
It must be noted again that due to the absence of suitable DPDFs to
NNLO accuracy, only DPDFs which have been determined to NLO accuracy
could be used for our predictions.
A consistent treatment of higher order contributions to the hard
matrix elements for all processes entering the fits of DPDFs will enable 
their consistent determination to NNLO. It 
 is considered to be of crucial
importance for future improvements for predictions of DDIS processes. 
\begin{table*}[t!bh]
    \caption{
      NNLO predictions for \HLRG using different DPDFs. 
      Mind, all DPDFs have been determined only in NLO accuracy.
      The uncertainties denote the DPDF uncertainty as provided by the
      respective DPDF sets.
    }
    \label{tab:DPDFs}
  \scriptsize
  \begin{center}
    \begin{tabular}{lcccccc}
            \hline
            {\bf Data set} & $\sigma^{\rm Data}$ 
            & $\sigma^{\text \DPDFFitA}$
            & $\sigma^{\text \DPDFFitB}$
            & $\sigma^{\text \DPDFFitJets}$
            & $\sigma^{\text \DPDFMRW}$
            & $\sigma^{\text \DPDFZSJ}$ \\
            & [pb] & [pb] & [pb] & [pb] & [pb]& [pb] \\
            \hline
            \HLRG & 
            $73\pm7_{\rm exp}$  
            & $129.3^{+16.8}_{-20.4}$  
            & $98.6^{+11.7}_{-14.7}$  
            & $83.1$  
            & $101.8$  
            & $78.0$   \\
      \hline
    \end{tabular}
    \end{center}
\end{table*}

\subsection{Differential distributions}
In total we computed 39 single-differential distributions and four double-differential distributions for 
available measurements, which are summarised in
table~\ref{tab:datasetsDistributions}.

\begin{table*}[tb]
    \caption{Overview of the measured single- and double-differential distributions.}
    \label{tab:datasetsDistributions}
  \scriptsize
  \begin{center}
    \begin{tabular}{c|cccccc}
        \hline
            Histogram         & H1 FPS  & H1 VFPS & H1 LRG  & H1 LRG &  H1 LRG   & ZEUS LRG     \\
                              & \HERAII & \HERAII & \HERAII & \HERAI &   \LowEP  & \HERAI   \\
            \hline                                                               
            $\Qsq$            &   \ok   &  \ok    &   \ok   &        &    \ok    &  \ok       \\
            $y ~[W]^\ast$     &   \ok   &  \ok    &   \ok   &   \ok  &    $\ast$ & $\ast$     \\           
       $\ptjone~[\ptjet]^\ast$&   \ok   &  \ok    &   \ok   &   \ok  &    \ok    &  $\ast$    \\ 
            $\ptavg$          &         &         &   \ok   &        &           &           \\
            $\ptjtwo$         &         &         &   \ok   &        &           &           \\
      $\etaavg ~[\etaj]^\ast$ &         &   \ok   &         &        &    \ok    & $\ast$   \\
 $\deleta ~[\deletastar]^\ast$&  $\ast$ &   \ok   &   $\ast$&  $\ast$&    $\ast$ &             \\
            $\mx $           &         &   \ok   &         &        &           & \ok          \\
            $\xpom$           &  \ok    &   \ok   &  \ok    &  \ok   &  \ok      &           \\
            $\zpom$           &  \ok    &   \ok   &  \ok    &  \ok   &           & \ok          \\ 
            \hline                                                               
            $(\Qsq;\ptjone)$  &         &         &  \ok    &        &           &          \\
            $(\Qsq;\zpom)$    &         &         &  \ok    &        &           &  \ok     \\
            $(\ptjone;\zpom)$ &         &         &         &        &           &  \ok      \\
      \hline
    \end{tabular}
    \end{center}
\end{table*}
%

\begin{figure*}[tbh]
\centering
  \includegraphicss[trim={2.1cm 0 1.8cm 0},clip,width=0.8\textwidth]{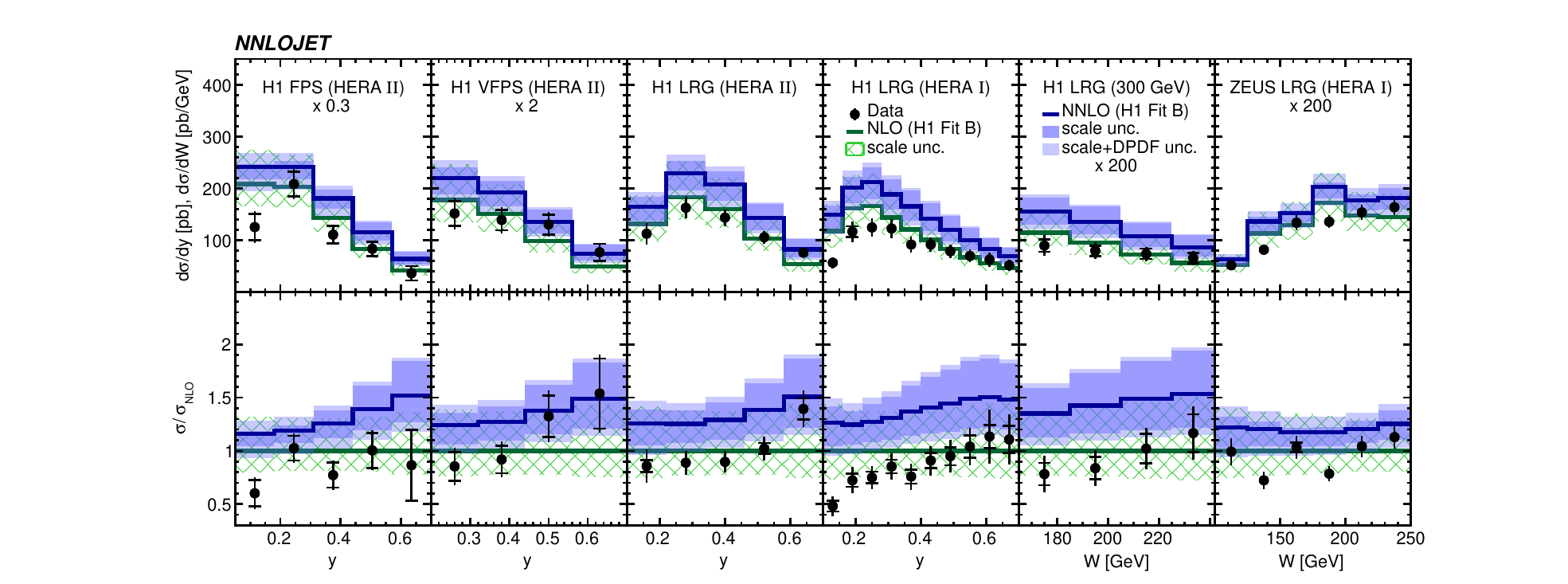} 
  \caption{
    The differential cross sections as a function of $y$ or,
    equivalently, $W$.
    In the upper panel, some of the distributions are scaled by a
    constant factor for better visibility.
    Displayed are the NNLO predictions in comparison to data and NLO
    predictions.
    The lower panel displays the ratio to NLO predictions.
    The shaded (hatched) area indicates the scale uncertainty of the
    NNLO (NLO) predictions.
    The bright shaded area around the NNLO predictions displays the
    scale and DPDF uncertainty added in quadrature.    
  }
  \label{fig:diffShape}
\end{figure*}

The NNLO predictions and their ratio to NLO predictions 
as a function of the inelasticity $y$
are displayed together with their experimental data in figure~\ref{fig:diffShape}.
The inelasticity $y$ is related to the $\gamma^*p$ centre-of-mass energy by
$W\simeq\sqrt{ys}$.
The NNLO predictions provide an improved description of the shape of
the data compared to respective NLO predictions, while being too high in their normalisation.
The NNLO scale uncertainty is significantly reduced in comparison to
the NLO scale uncertainty, which is most distinct at lower values of $y$.

\begin{figure*}[tbp]
\centering
\includegraphicss[trim={2.1cm 0 1.8cm 0},clip,width=0.79\textwidth]{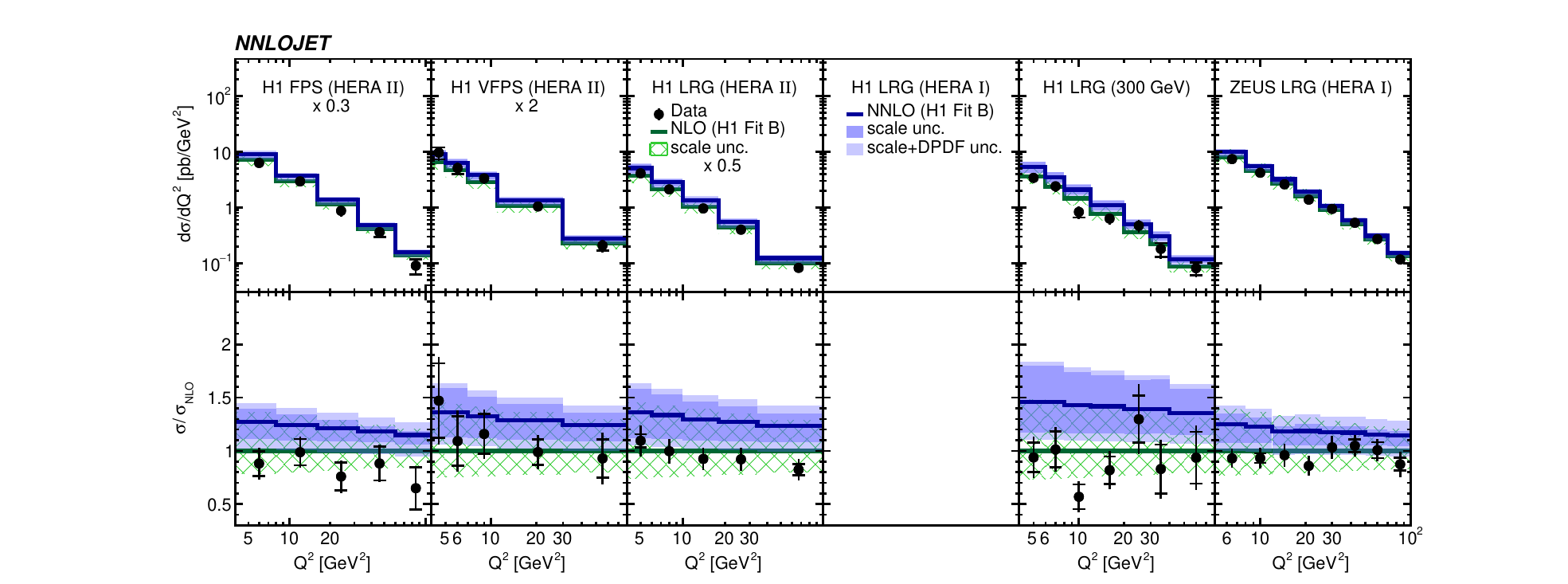} 
\caption{
  The differential cross sections as a function of \Qsq.
  In case where the panel is empty, the respective analysis did not
  provide a measurement of the displayed observable. 
  Other details as in figure~\ref{fig:diffShape}.
  \label{fig:Q2}
}
\end{figure*}

\begin{figure*}[tbp]
\centering
\includegraphicss[trim={2.1cm 0 1.8cm 0},clip,width=0.79\textwidth]{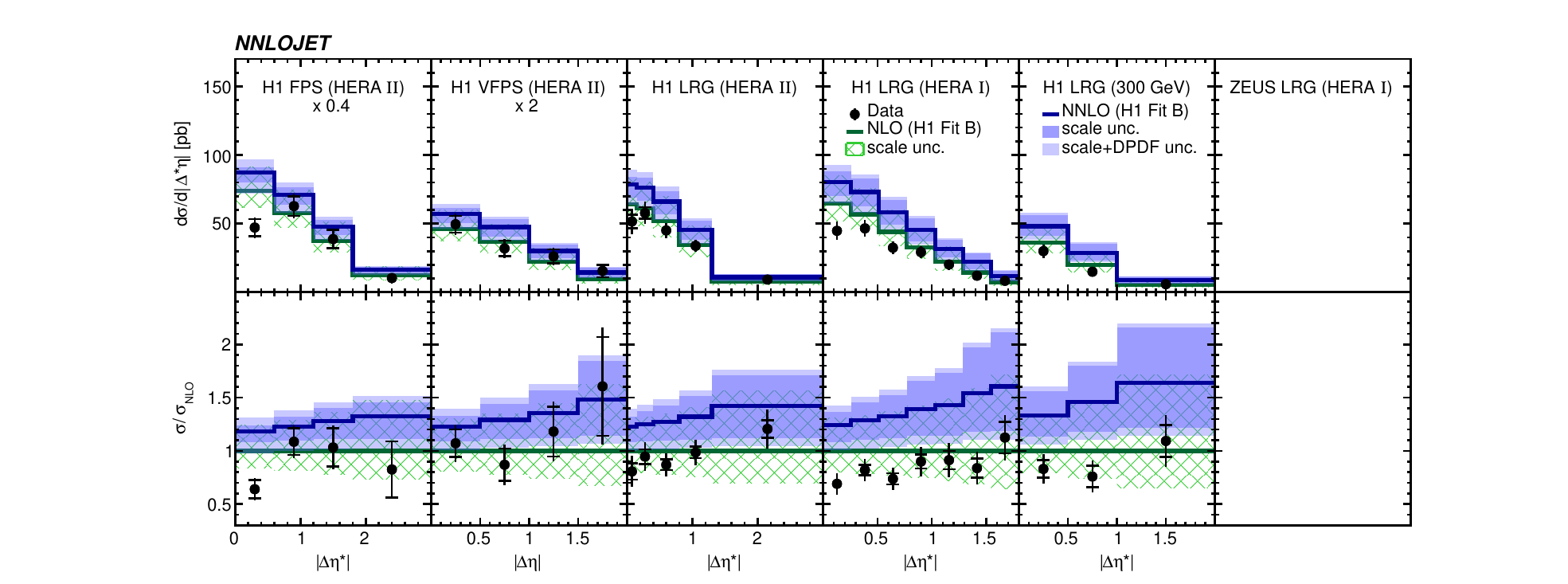}
\caption{
      The differential cross sections as a function of
      $|\deletastar|$ or $|\Delta \eta|$.
      Other details as in figure~\ref{fig:diffShape}.
}
\label{fig:deleta}
\end{figure*}

\begin{figure*}[tbp]
\centering
  \includegraphicss[trim={2.1cm 0 1.8cm 0},clip,width=0.79\textwidth]{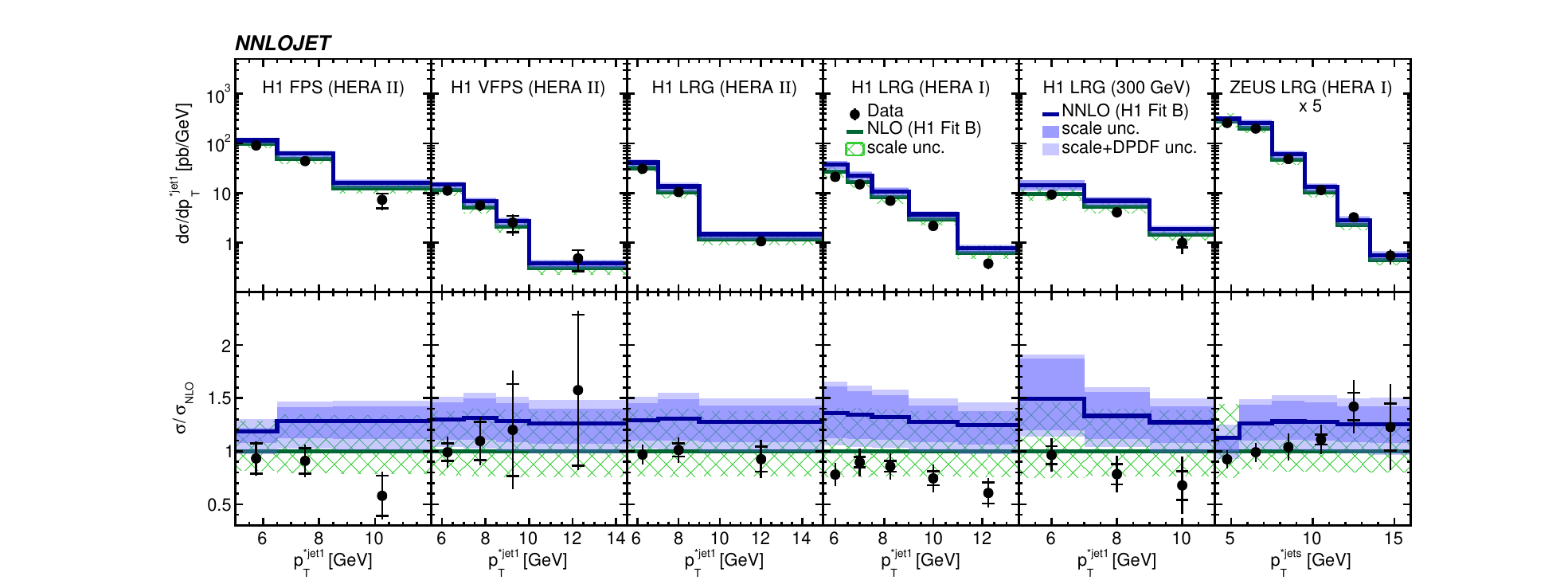} 
\caption{
  The differential cross sections as a function of \ptjone\ or \ptjet.
  Other details as in figure~\ref{fig:diffShape}.
  \label{fig:diffPt}
}
\end{figure*}

\begin{figure*}[tbp]
\centering
\includegraphicss[trim={2.1cm 0 11.3cm 0},clip,height=6.0cm]{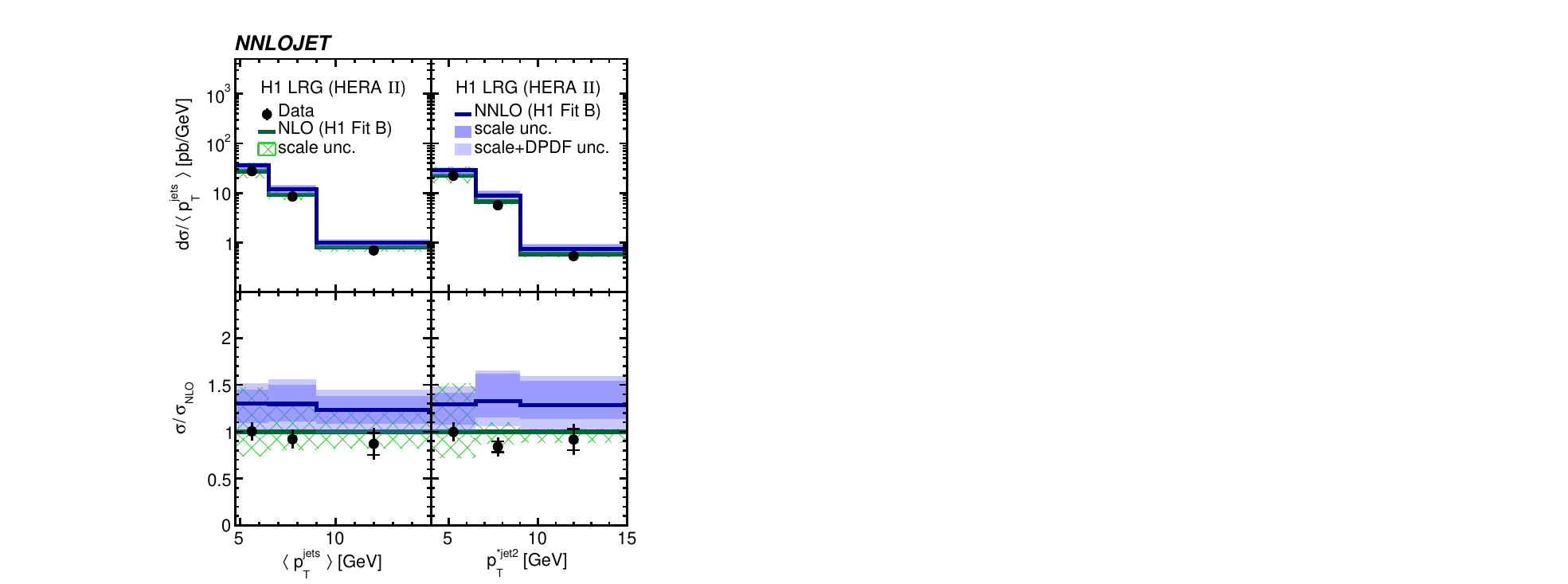}
\includegraphicss[trim={2.1cm 0 11.3cm 0},clip,height=6.0cm]{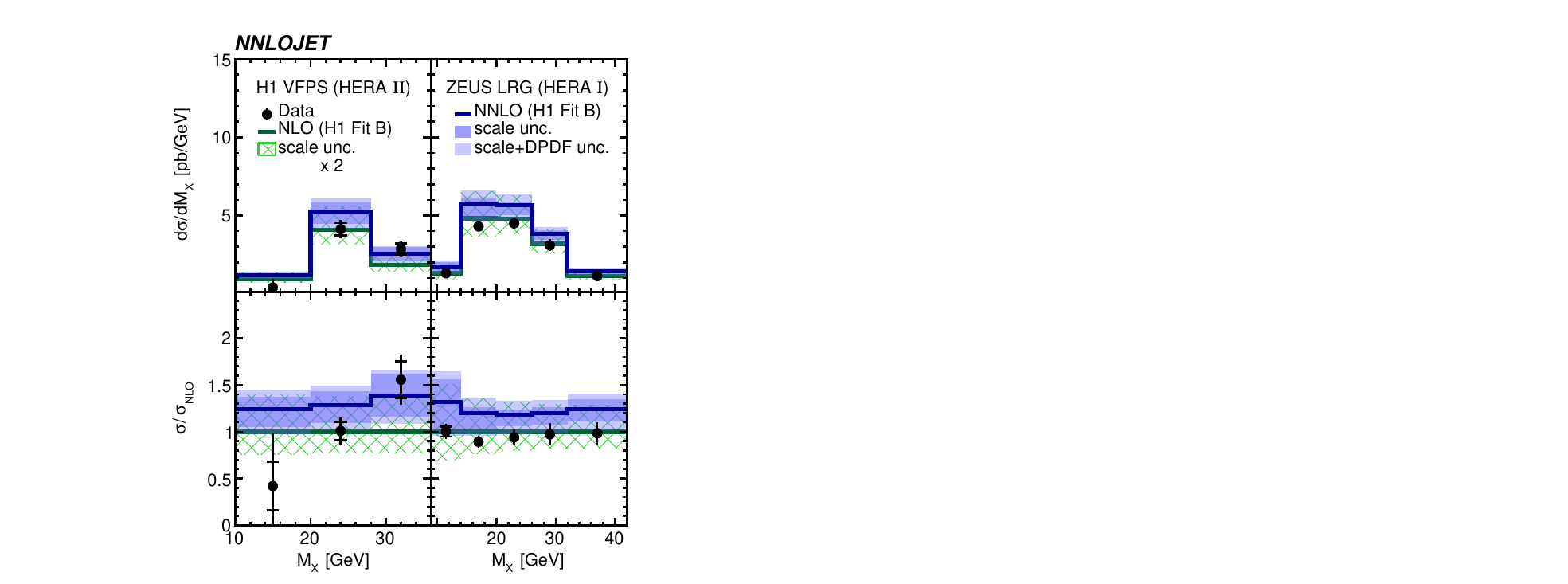}
\caption{
  The differential cross sections
  as a function of $\ptavg$ and $\ptjtwo$ as measured in \HLRG (left), and
  as a function of $\mx$  as measured in \HVFPS and \ZLRG (right).
  Other details as in figure~\ref{fig:diffShape}.
 }
  \label{fig:ptAvgMx}
\end{figure*}

\begin{figure*}[tbp]
\centering
\includegraphicss[trim={2.1cm 0 1.8cm 0},clip,width=0.79\textwidth]{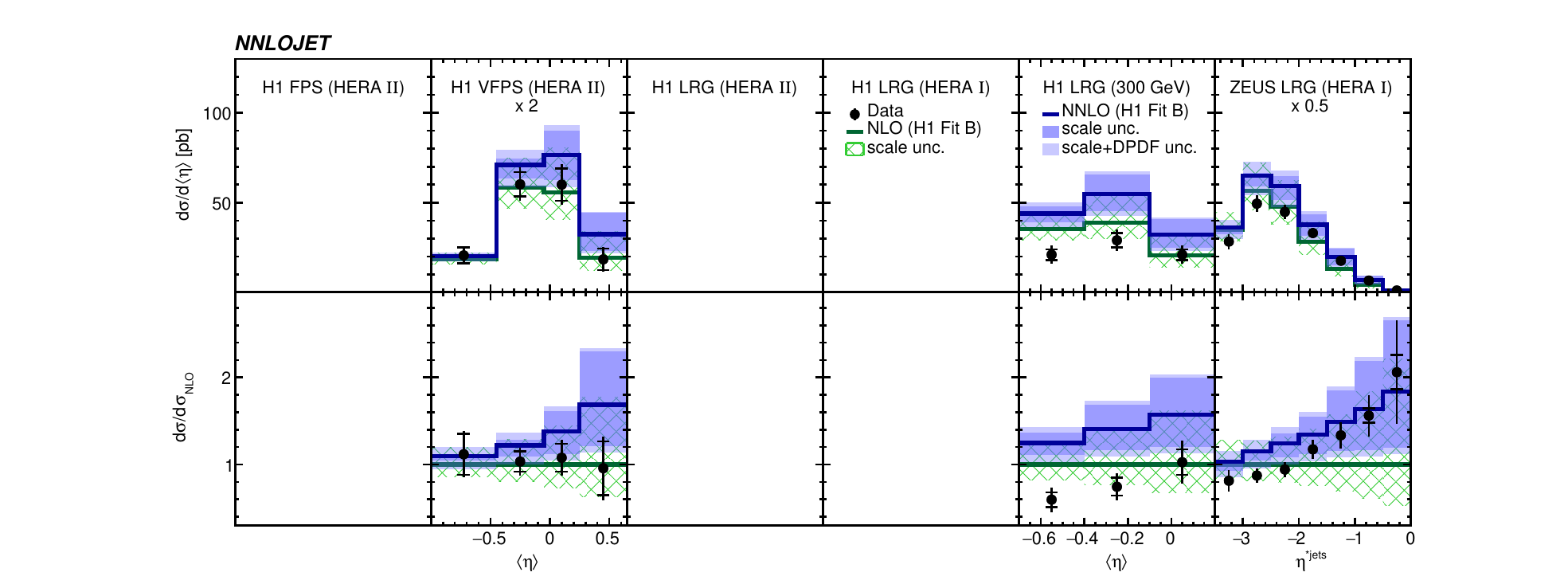} 
\caption{
  The differential cross sections as a function of $\etaavg$ or $\etaj$.
  Other details as in figure~\ref{fig:diffShape}.
}
  \label{fig:etaJets}
\end{figure*}

\begin{figure*}[tpb]
\centering
\includegraphicss[trim={2.1cm 0 1.8cm 0},clip,width=0.79\textwidth]{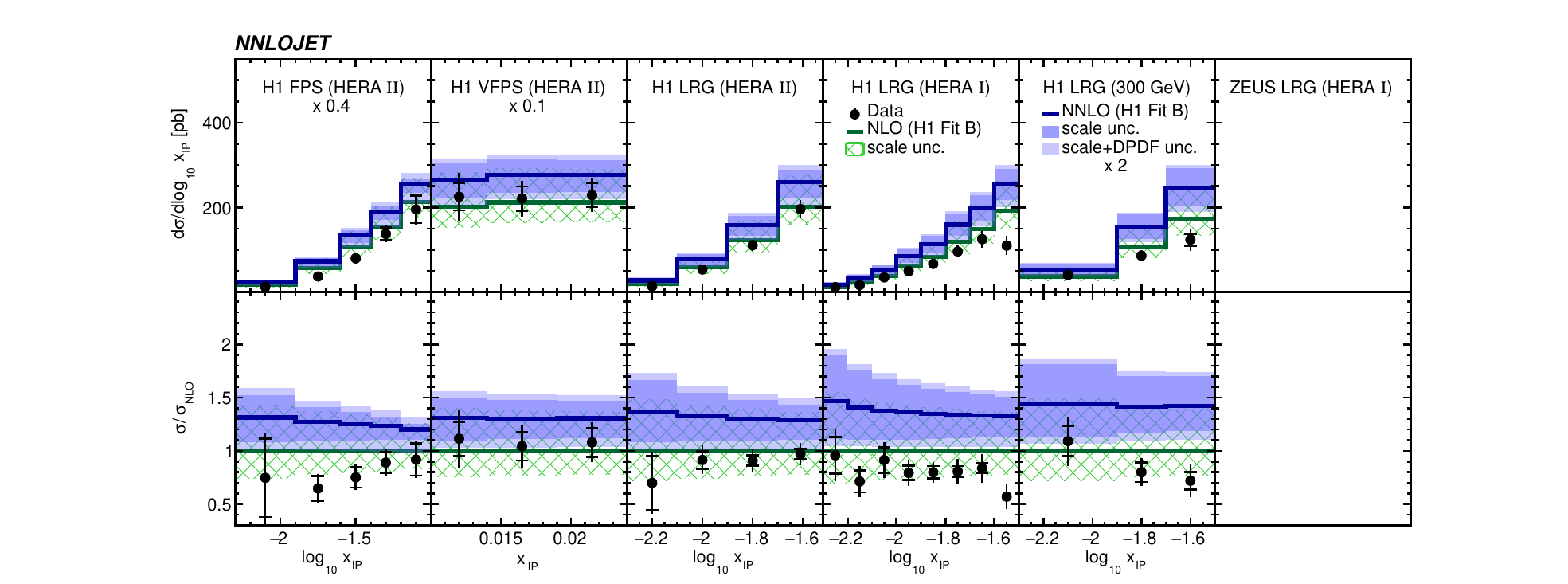} 
\caption{
  The differential cross sections as a function of \xpom.
  Other details as in figure~\ref{fig:diffShape}.
}
  \label{fig:xpom}
\end{figure*}

\begin{figure*}[tbp]
\centering
\includegraphicss[trim={2.1cm 0 1.8cm 0},clip,width=0.79\textwidth]{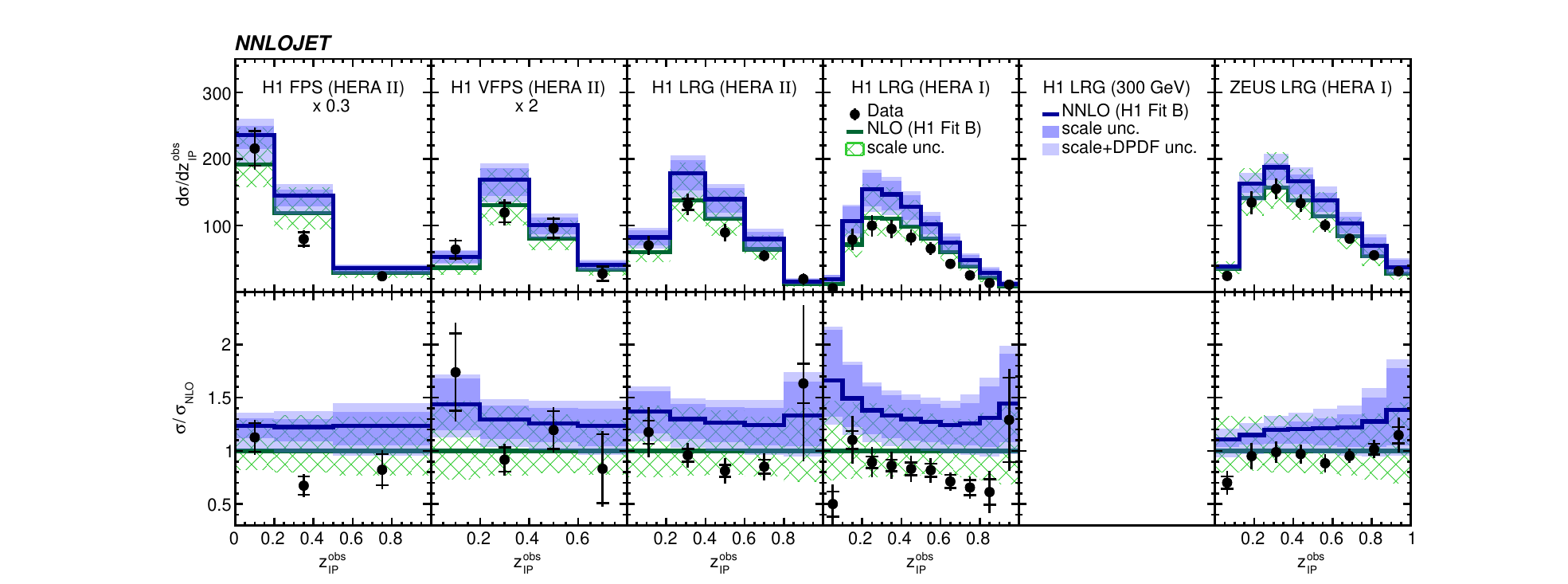} 
\caption{
  The differential cross sections as a function of \zpom.
  Other details as in figure~\ref{fig:diffShape}.
}
  \label{fig:zpom}
\end{figure*}

\begin{figure*}[tbp]
\centering
\includegraphicss[trim={2.1cm 0 6.4cm 0},clip,height=6.0cm]{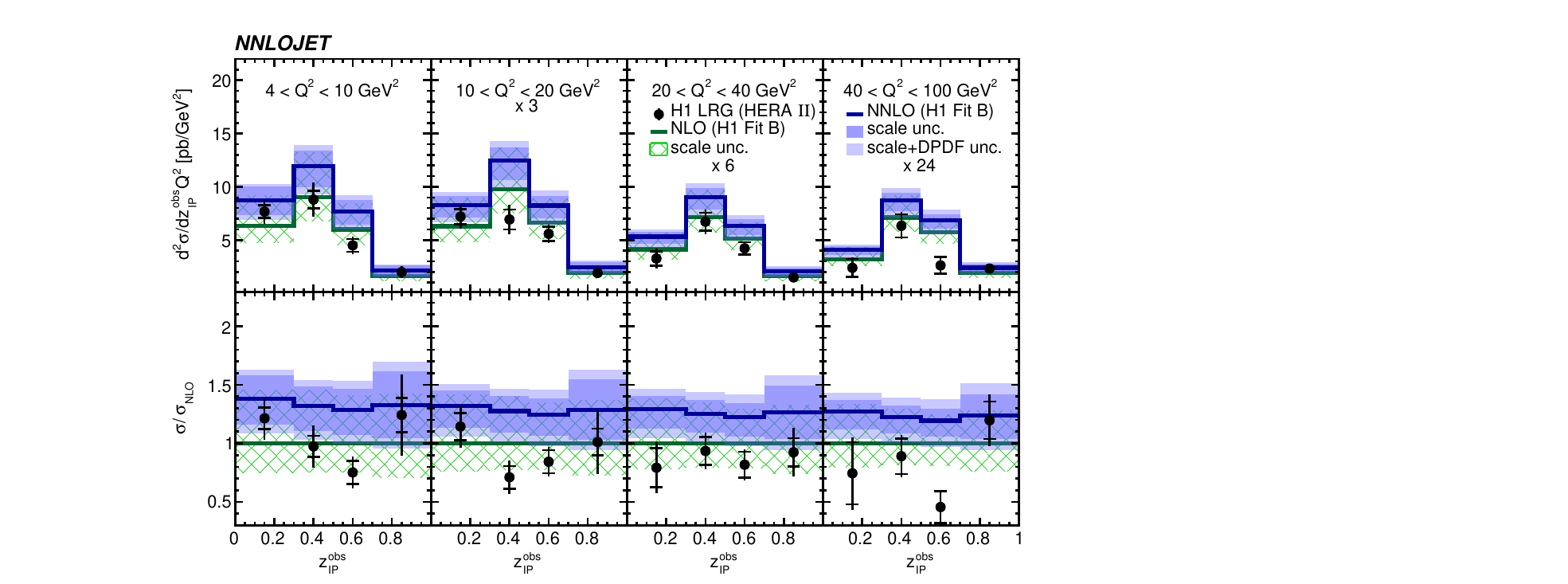}
\caption{The double-differential cross sections as functions of $\zpom$ and $Q^2$ as measured in \HLRG.
  Other details as in figure~\ref{fig:diffShape}.
 }
\label{fig:H1LRG2DzpomQ2}
\end{figure*}

\begin{figure*}[tbp]
\centering
\includegraphicss[trim={2.1cm 0 6.4cm 0},clip,height=6.0cm]{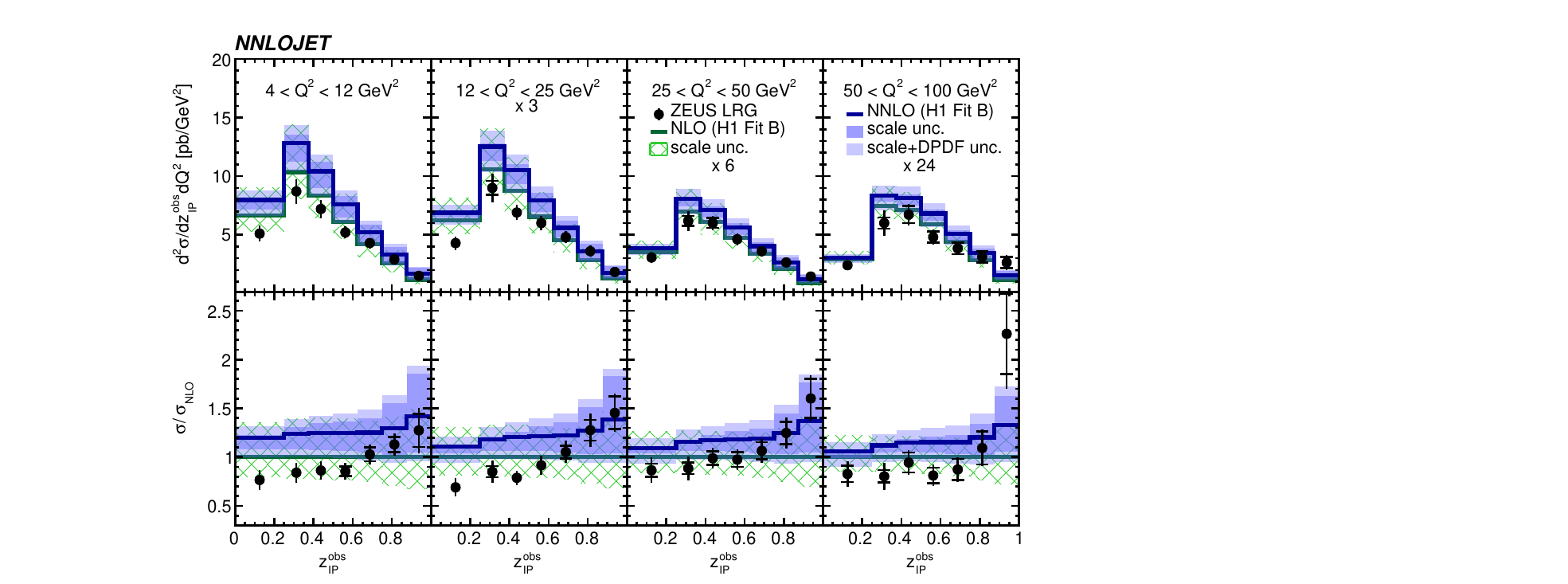}
\caption{The double-differential cross sections as functions of $\zpom$ and $Q^2$ as measured in \ZLRG.
  Other details as in figure~\ref{fig:diffShape}.
 }
\label{fig:ZEUSzpomQ2}
\end{figure*}

\begin{figure*}[tbp]
\centering
\includegraphicss[trim={2.1cm 0 3.9cm 0},clip,height=6.0cm]{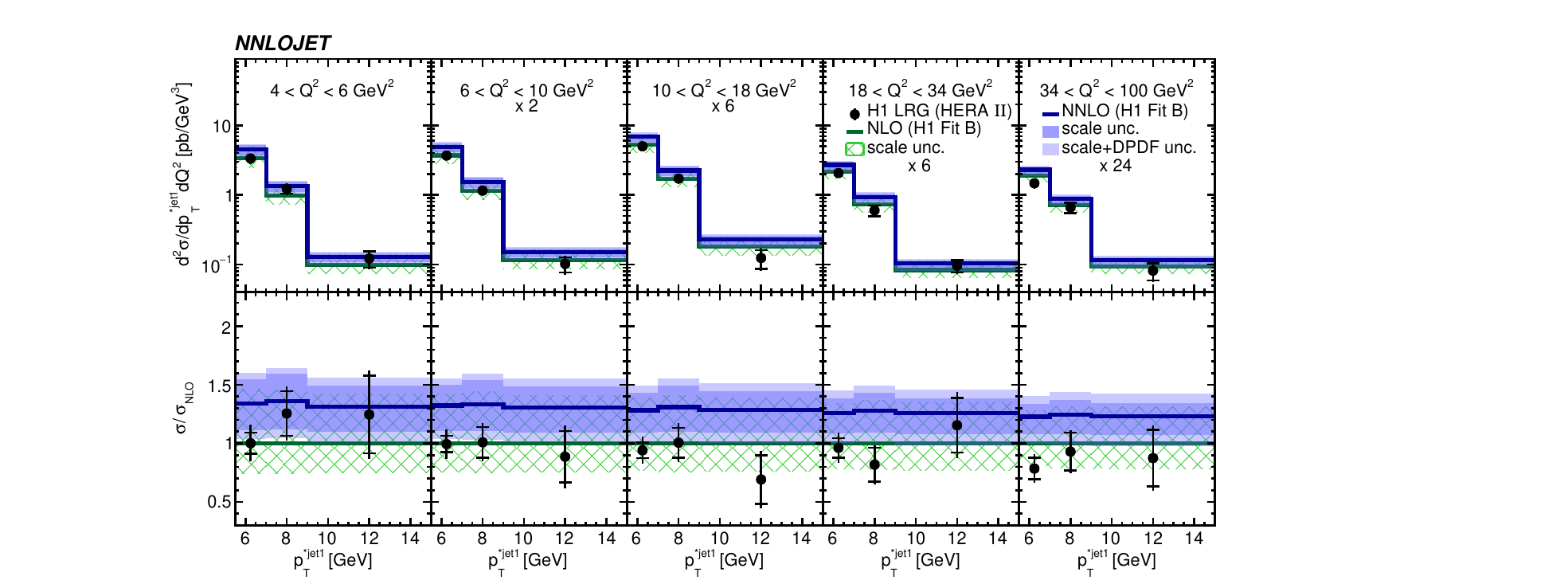}
\caption{
  The double-differential cross sections as functions of $\ptjone$ and $Q^2$ as measured in \HLRG.
  Other details as in figure~\ref{fig:diffShape}.
 }
\label{fig:H1LRG2DptQ2}
\end{figure*}

\begin{figure*}[tbp]
\centering
\includegraphicss[trim={2.1cm 0 8.8cm 0},clip,height=6.0cm]{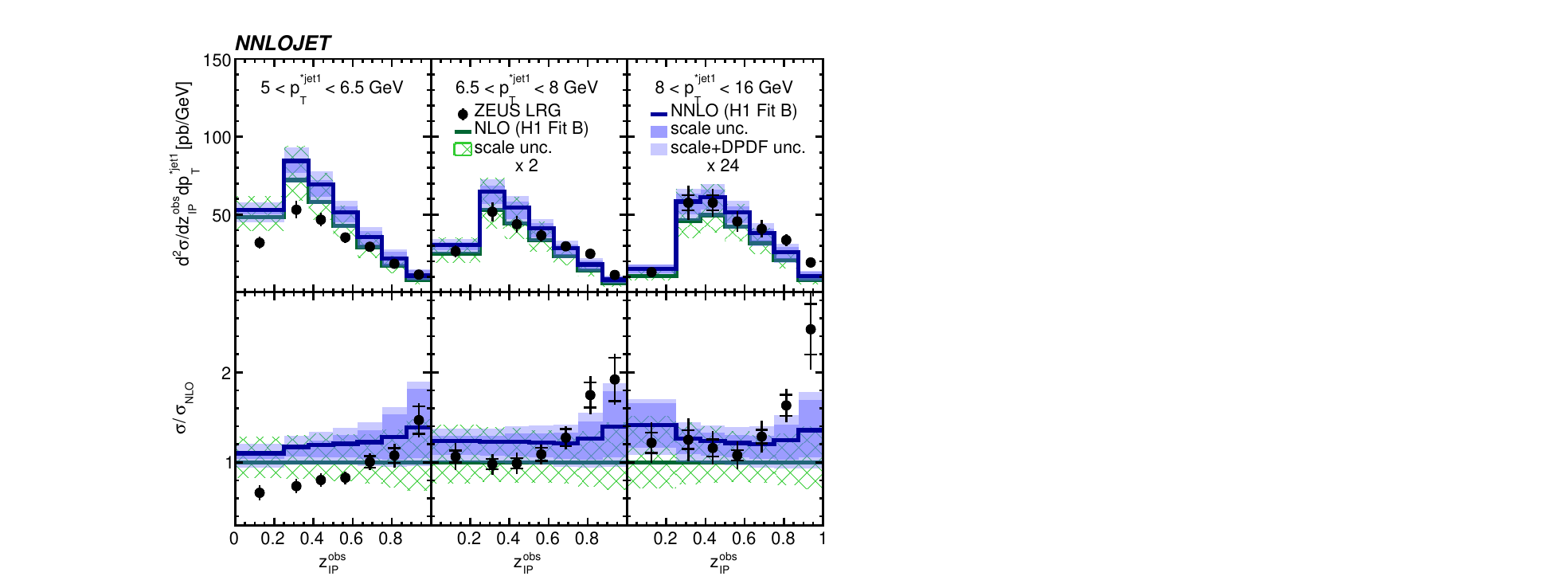}
\caption{The double-differential cross sections as functions of $\zpom$ and $\ptjone$ as measured in \ZLRG.
  Other details as in figure~\ref{fig:diffShape}.
 }
\label{fig:ZEUSzpomPtone}
\end{figure*}

The NNLO predictions as a function of \Qsq, $|\deletastar|$ (or
$|\Delta \eta|$), \ptjone\ (or \ptjet), $\ptavg$, $\ptjtwo$, $\mx$,
$\etaavg$ (or  $\etaj$),  \xpom\ and \zpom\ are presented in
figures~\ref{fig:Q2} to~\ref{fig:zpom}, respectively, and compared to data.
Double-differential predictions as functions of \zpom
and \ptjone for \Qsq  intervals, and
as a function of \zpom for \ptjone  intervals are presented in
figures~\ref{fig:H1LRG2DzpomQ2} to~\ref{fig:ZEUSzpomPtone}.
Similar conclusions as for the $y$ distribution can be drawn from
these comparisons.
Some variants of selected distributions are discussed in more detail
in the following.
%

While $y$ is an inclusive observable, the rapidity separation of the
two leading jets, $|\deletastar|$, is directly sensitive to effects
 emerging from higher order radiative corrections.
Also for this observable, the NNLO predictions provide an improved
description of the shape for measured distributions, as can be seen
in figure~\ref{fig:deleta}.
Similar observations are made for all remaining distributions. 
This in particular for distributions in \Qsq, $\langle\eta\rangle$ and \zpom~(see
figures~\ref{fig:Q2},~\ref{fig:etaJets} and~\ref{fig:zpom}).

\begin{figure*}[tbh]
\centering
  \includegraphicss[trim={2.1cm 0 1.8cm 0},clip,width=0.79\textwidth]{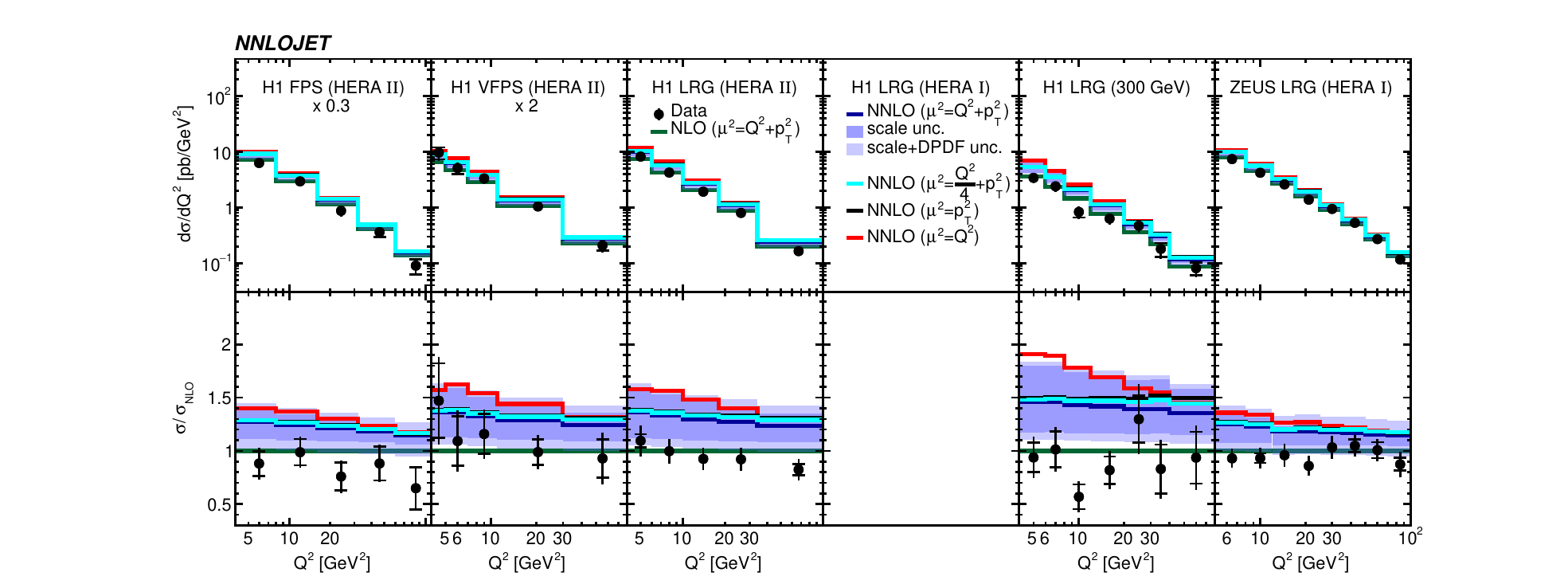} 
  \caption{
    The differential cross sections as a function of \Qsq.
    Displayed are NNLO predictions for different scale definitions.
    Further details are given in figure~\ref{fig:diffShape}.
  }
  \label{fig:diffScales}
\end{figure*}

NNLO predictions as a function of \Qsq\
obtained with different scale definitions
are displayed in figure~\ref{fig:diffScales}.
For this study we set $\mu := \mu_F = \mu_R$.
The studied scale definitions 
$\mu^2 = \Qsq/4 + \ptavg^2$ and 
$\mu^2 = \ptavg^2$ provide similar results as the nominal scale
definition of $\mu^2 = Q^2 + \ptavg^2$, whereas the
scale choice $\mu^2 = \Qsq$ results in higher cross sections and a
steeper \Qsq\ spectrum. 
The studied scale choices are covered by the scale uncertainties.

\begin{figure*}[tbhp]
\centering
  \includegraphicss[trim={2.1cm 0 1.8cm 0},clip,width=0.8\textwidth]{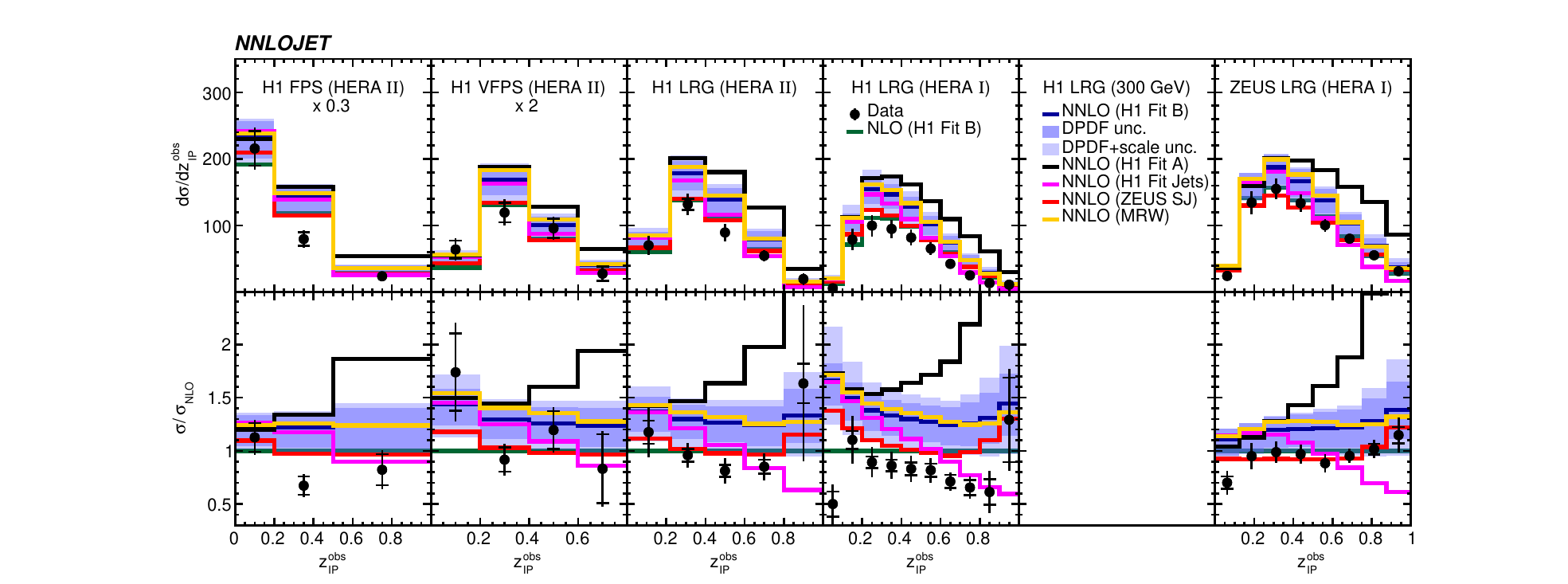}
  \caption{
    The differential cross sections as a function of \zpom.
    Displayed are NNLO predictions for different DPDFs.    
    Further details are given in figure~\ref{fig:diffShape}.
  }
  \label{fig:diffDPDFs}
\end{figure*}
NNLO predictions for \zpom distributions obtained using different DPDFs
are displayed in figure~\ref{fig:diffDPDFs}.
For this observable, NNLO predictions using the \DPDFFitB\ and \DPDFMRW\ DPDFs give quite similar
results and lie above most of the data.
Results obtained with the \DPDFFitA DPDF significantly overestimate the measurements
in particular for higher values of \zpom.
Predictions obtained with \DPDFZSJ\ and \DPDFFitJets\ give lower cross sections, 
but the application of the \DPDFFitJets\ DPDF also results in a considerably different shape of the distribution.
In general, the latter two DPDFs, including dijet data in their determination, give an improved description of the data compared to the first two DPDFs.
It should be noted however, that differences arising from applications of different DPDFs are not covered by the uncertainties taken from the \DPDFFitB\ DPDF.
This feature is most prominent at higher values of \zpom.

\begin{figure*}[ht]
\centering
  \includegraphicss[trim={2.1cm 0 1.8cm 0},clip,width=0.79\textwidth]{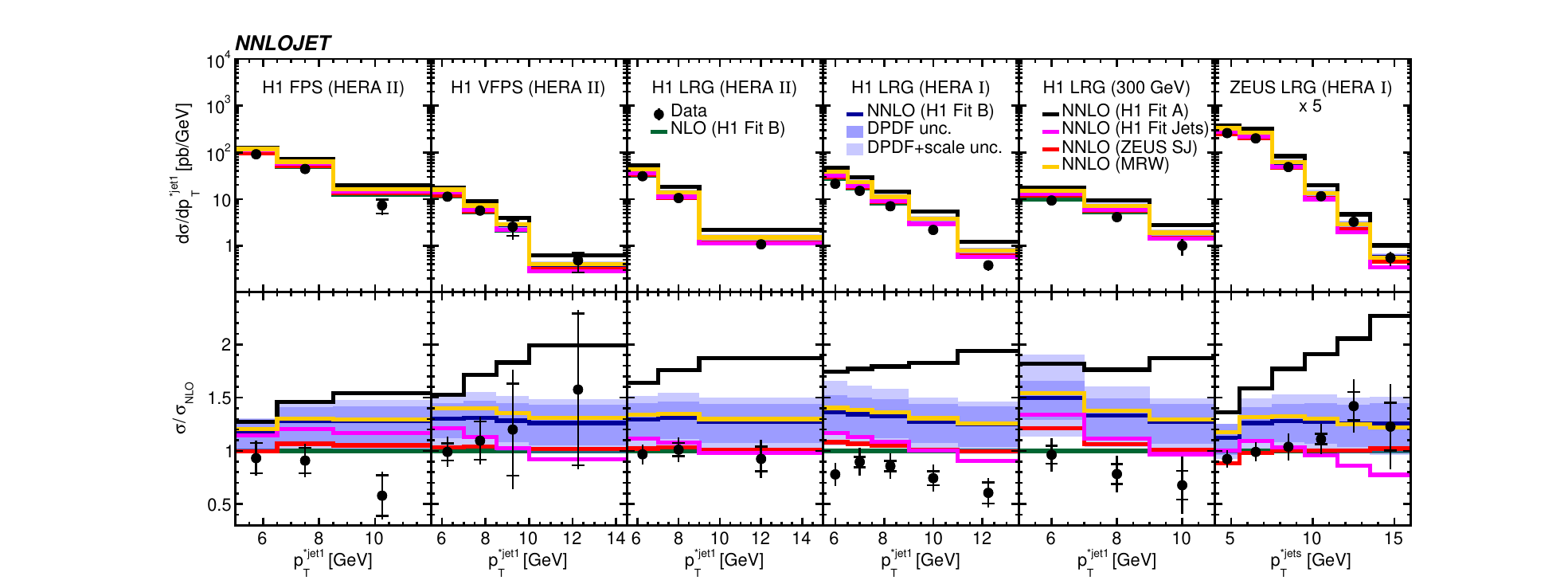}
\caption{
  The differential cross sections as a function of \ptjone\ or \ptjet obtained for different DPDFs.    
  Other details as in figure~\ref{fig:diffShape}.
  \label{fig:diffDPDFsPt}
}
\end{figure*}

In summary, NNLO predictions using the stated DPDFs provide an overall satisfactorily
description of the data.
However, none of the studied DPDFs is able to describe the
shapes of the distributions of all of the \ptjone (or \ptjet)
measurements equally well, as can be seen from their  
comparisons to predictions displayed in figure~\ref{fig:diffDPDFsPt}.

The studied DPDFs mainly differ in their gluon component~\cite{Zlebcik:2011kq}.
This explains the observed differences between results obtained with
 different DPDFs as the gluon is the most important parton inside the DPDFs.
It is therefore crucial to determine the gluonic component of the DPDFs
more accurately, and once this is achieved, theoretical predictions are expected to provide an improved description of the data.

\begin{figure*}[tbhp]
  \centering
  \includegraphicss[width=0.30\textwidth]{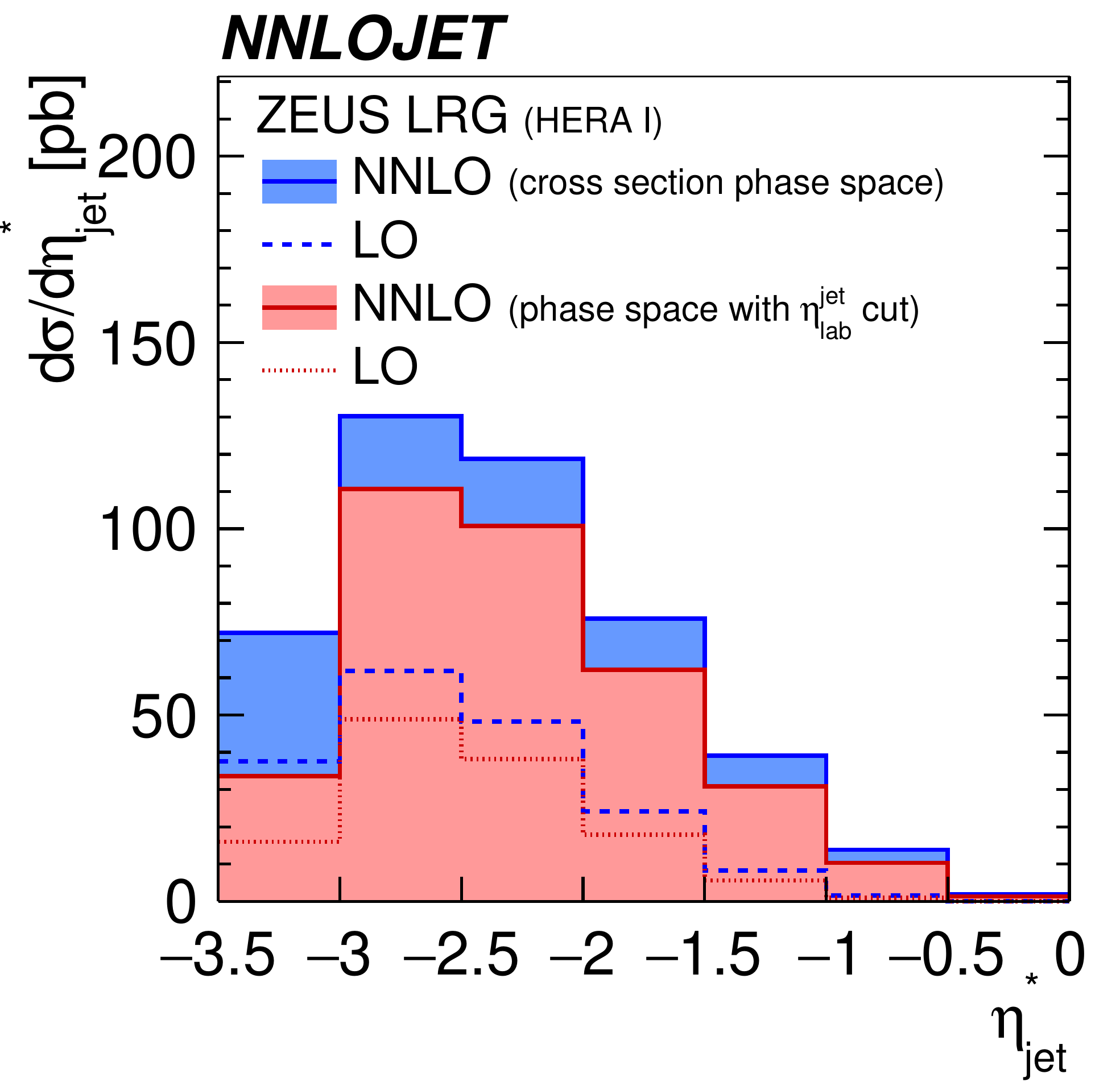}
  \includegraphicss[width=0.30\textwidth]{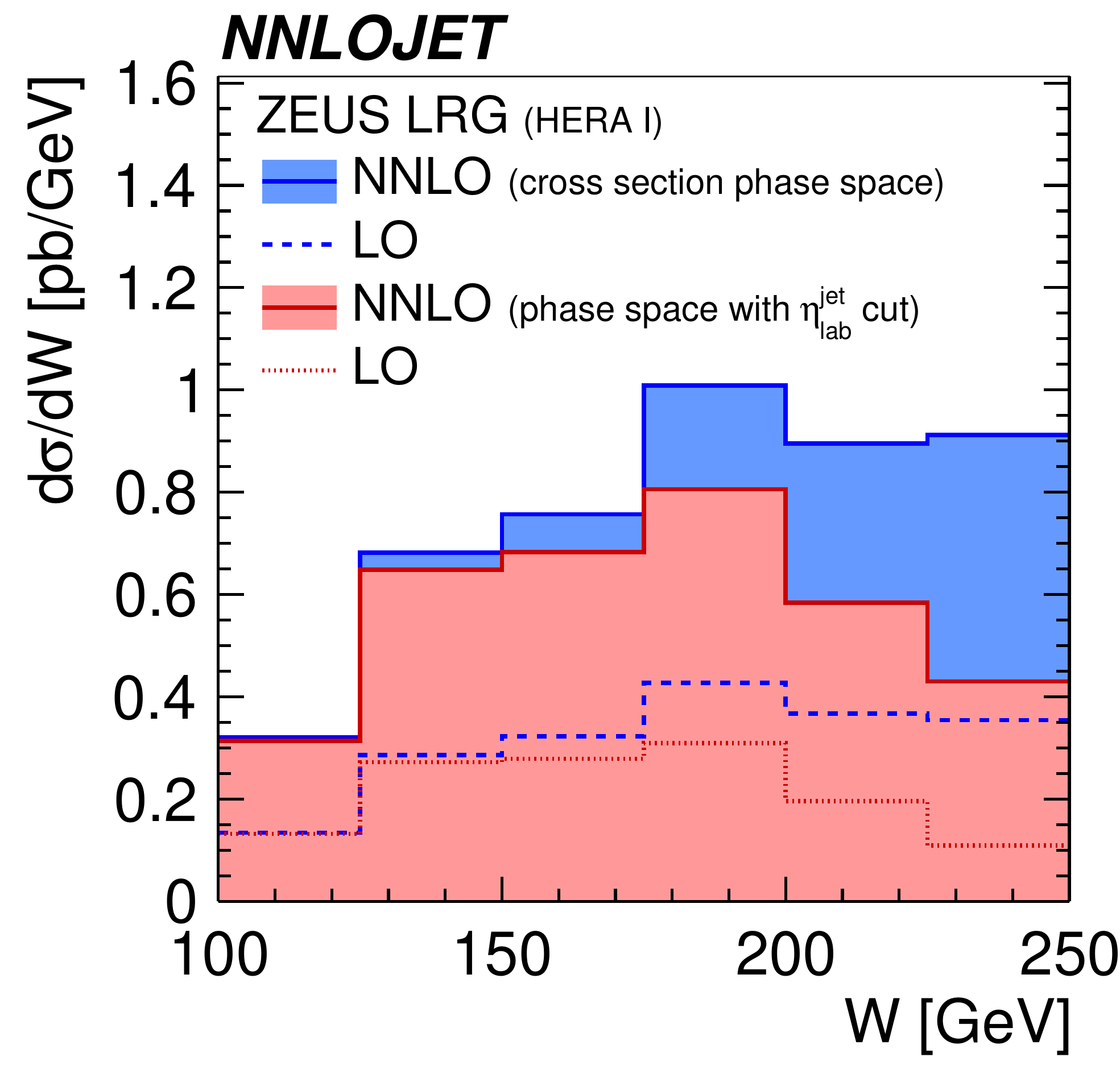}
  \includegraphicss[width=0.30\textwidth]{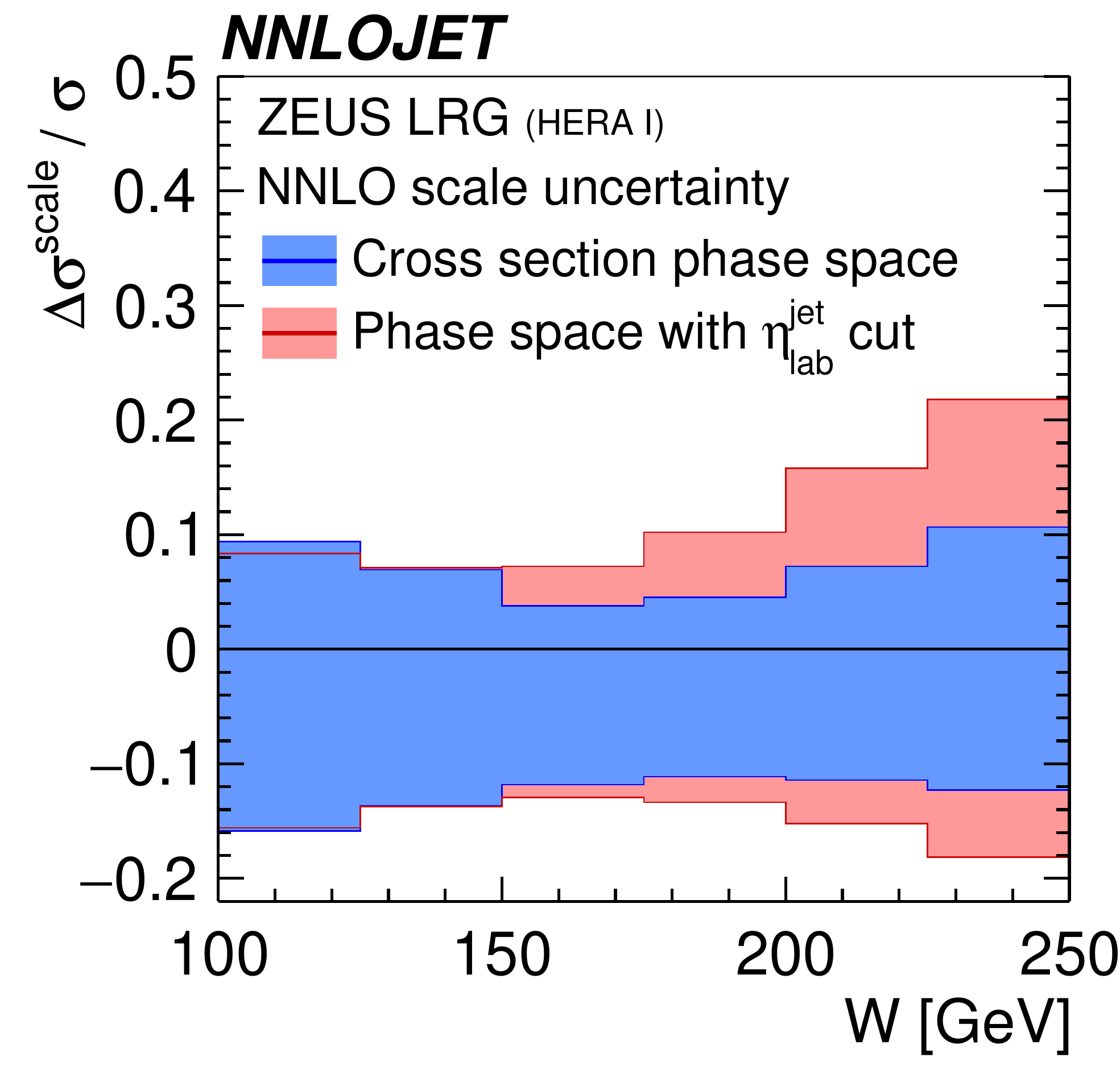}
  \caption{
    NNLO and LO predictions for the \ZLRG\ phase space with and
    without the
    additional cut of $-1<\etalab<2.5$ for two selected observables:
    $\etaj$ (left) and $W$ (middle).
    The right panel displays the relative NNLO scale uncertainty for
    the $W$ distribution for the two studied phase space definitions.
  }
  \label{fig:ZEUSps}
\end{figure*}

Despite the fact that the H1 and ZEUS experimental devices have a similar
resolution and comparable acceptances, it is observed that
predictions for the \ZLRG phase space often yield smaller scale
uncertainties as those for the comparable \HLRG\ phase space.
This is mainly due to the restriction on \etalab\ imposed by H1,
whereas  the ZEUS phase space is restricted only in $\etaj$,
even though an equivalent requirement on \etalab\ is imposed for
\ZLRG\ measurement on detector level~\cite{Chekanov:2007aa}.
In figure~\ref{fig:ZEUSps} a study is presented, where an additional \etalab\ cut of
$-1<\etalab<2.5$ on the NNLO and LO predictions for the \ZLRG phase
space is shown\footnote{The \ZLRG analysis
  required two jets to be within $-2<\etalab<2$~\cite{Chekanov:2007aa}. For better
  comparability and also due to technical reasons, we study an additional cut of
  $-1<\etalab<2.5$ in analogy to the \HFPS and \HVFPS measurements.}.
In particular at lower values of $|\etaj|$ and at higher values of
$W$, this cut would significantly reduce the cross section.
Once the additional cut on \etalab\ is imposed, the relative NNLO
scale uncertainty increases significantly, i.e.\ up to a factor of two in 
some parts of the phase space.
This becomes in particular distinct at high values of $W$, as
displayed in figure~\ref{fig:ZEUSps} (right).
In conclusion, it is observed that the phase space definition of \ZLRG
results in more stable pQCD predictions, i.e.\ lower scale
uncertainties, while important regions of the phase space
were not accessible by the experimental device and the extrapolation
factors were obtained by MC simulations.
Similar considerations also apply to the \HLRGI\ measurement.

\subsection{The gluon induced fraction}
\begin{figure}[tb]
  \centering
  \includegraphicss[width=0.32\textwidth]{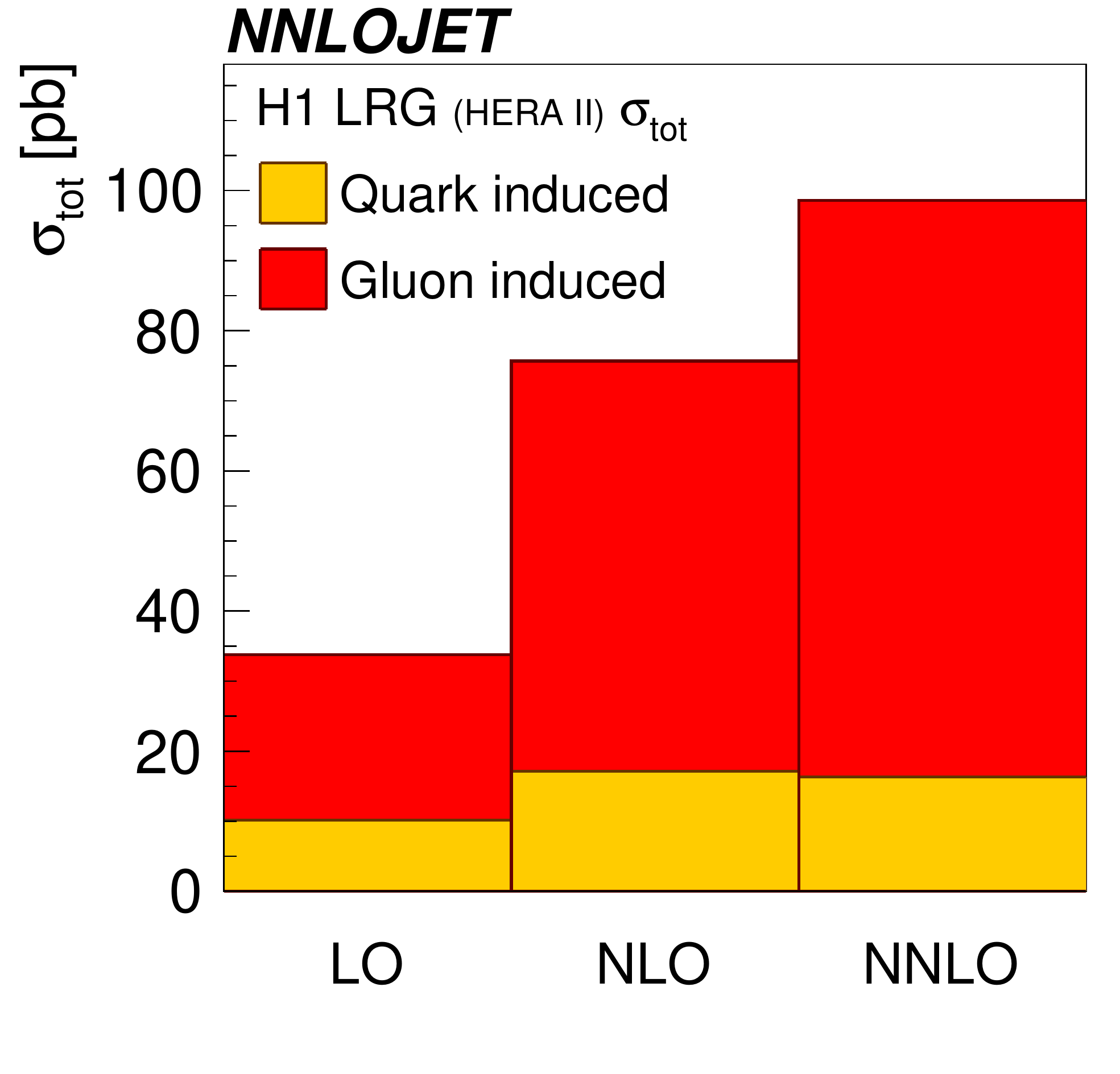}
  \caption{
  The decomposition of the \HLRG total dijet cross section into the part induced by gluons (red) and quarks (yellow).
  It is shown at LO, NLO and NNLO.  }
\label{fig:gluonPartTotal}
\end{figure}

In order to further elucidate the dependence of the NNLO predictions on the
individual parton flavors inside the DPDFs, 
the decomposition of the total \HLRG\ cross section
into gluon-induced and
quark-induced channels is shown for LO
, NLO and NNLO predictions in figure~\ref{fig:gluonPartTotal}.
It is apparent that the rise of the cross section at higher orders
is predominantly driven by the gluon-induced channels.

\begin{figure*}[tbhp]
  \centering
  \includegraphicss[width=0.32\textwidth]{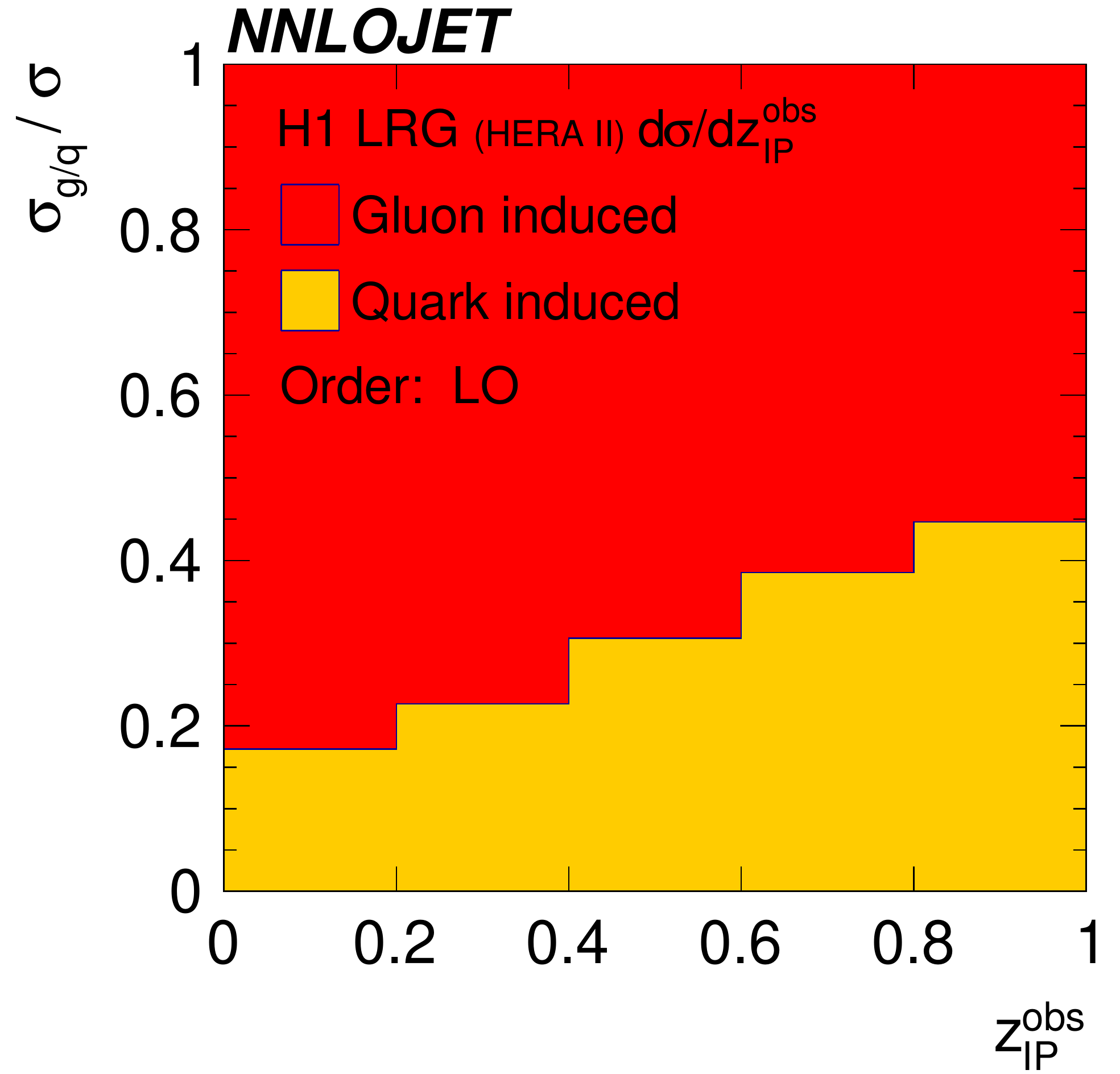}
  \includegraphicss[width=0.32\textwidth]{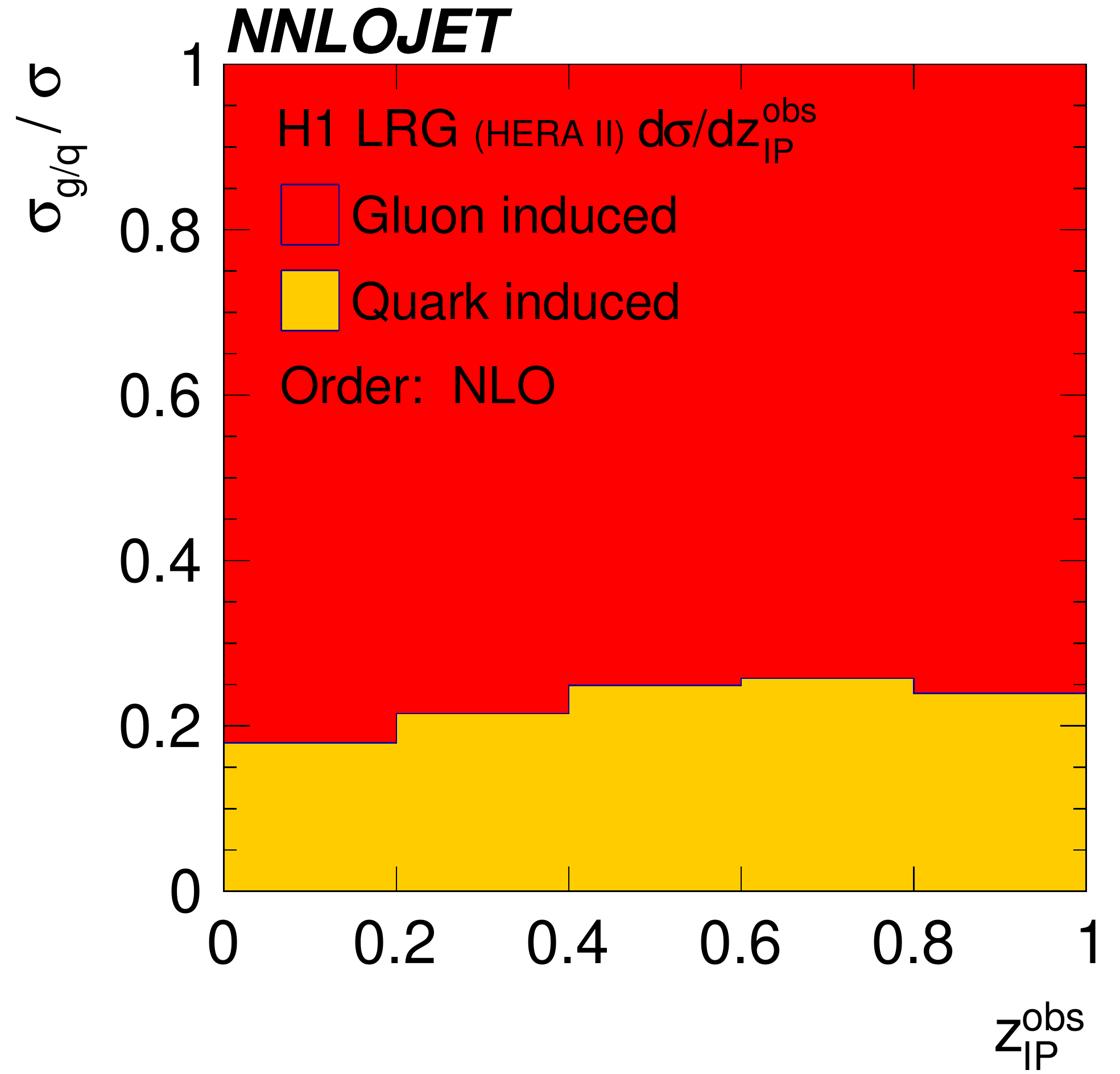}
  \includegraphicss[width=0.32\textwidth]{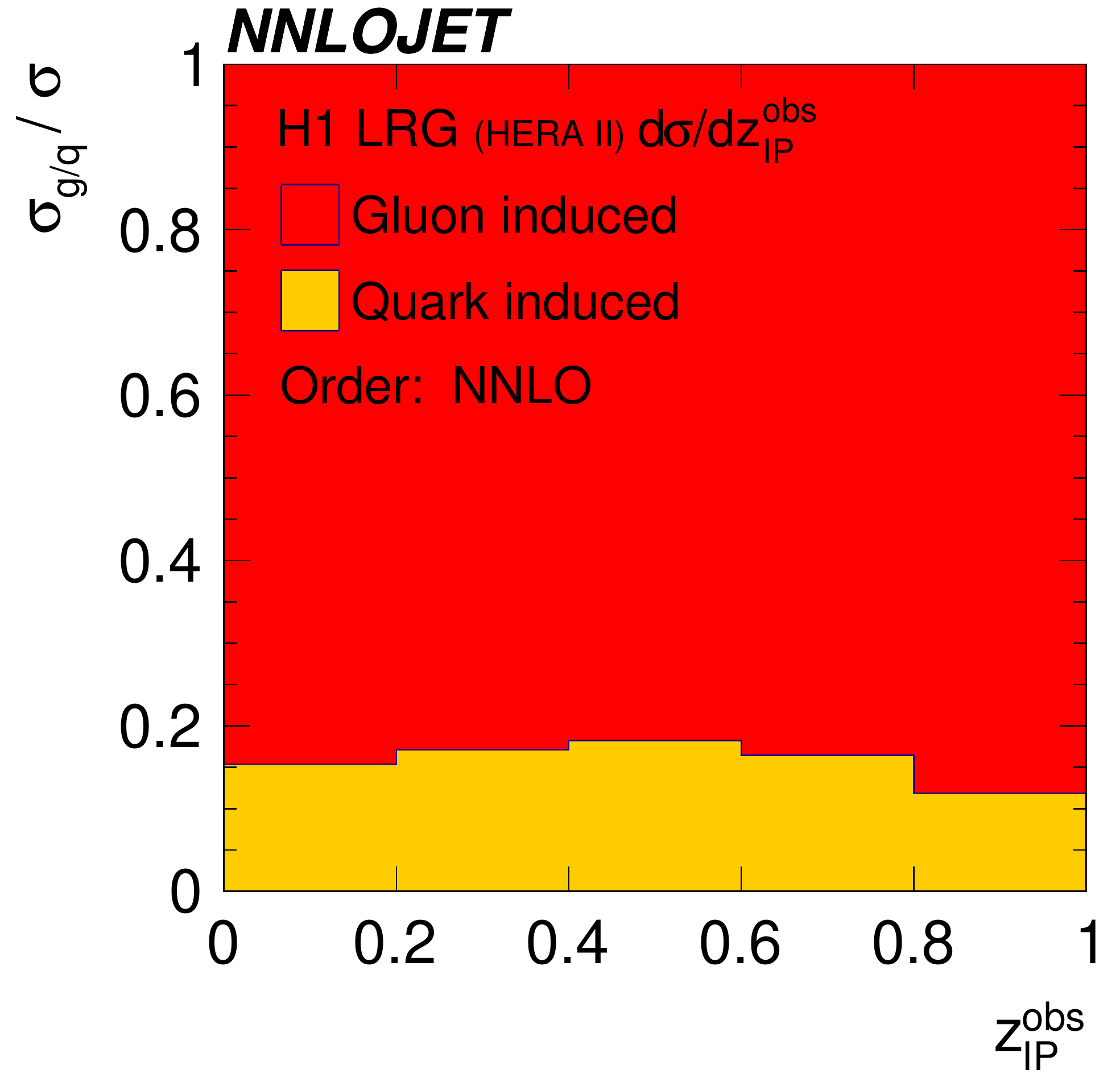}
  \caption{
  The fraction of gluon induced (red) and quark induced (yellow) contributions
  of the LO (left), NLO (middle) and NNLO (right) cross section as a
  function of \zpom. The kinematic range is adapted from the \HLRG
  measurement.}
\label{fig:gluonfraction}
\end{figure*}

The fractions of gluon- and quark-induced contributions to the
cross sections as a function of \zpom\ are displayed in
figure~\ref{fig:gluonfraction}.
While the fraction of the gluon-induced contribution remains unchanged for
different orders in $\alpha_s$ at low values of \zpom, there is a strong increase of
the gluon-induced fraction at higher values of \zpom for higher orders in $\alpha_s$.
Hence it can be deduced that future NNLO DPDFs are required to have a significantly reduced
gluon component as compared to currently available NLO DPDFs.
%

\subsection{The sensitivity to DPDFs}
\begin{figure*}[tb]
  \centering
  \includegraphicss[width=0.32\textwidth]{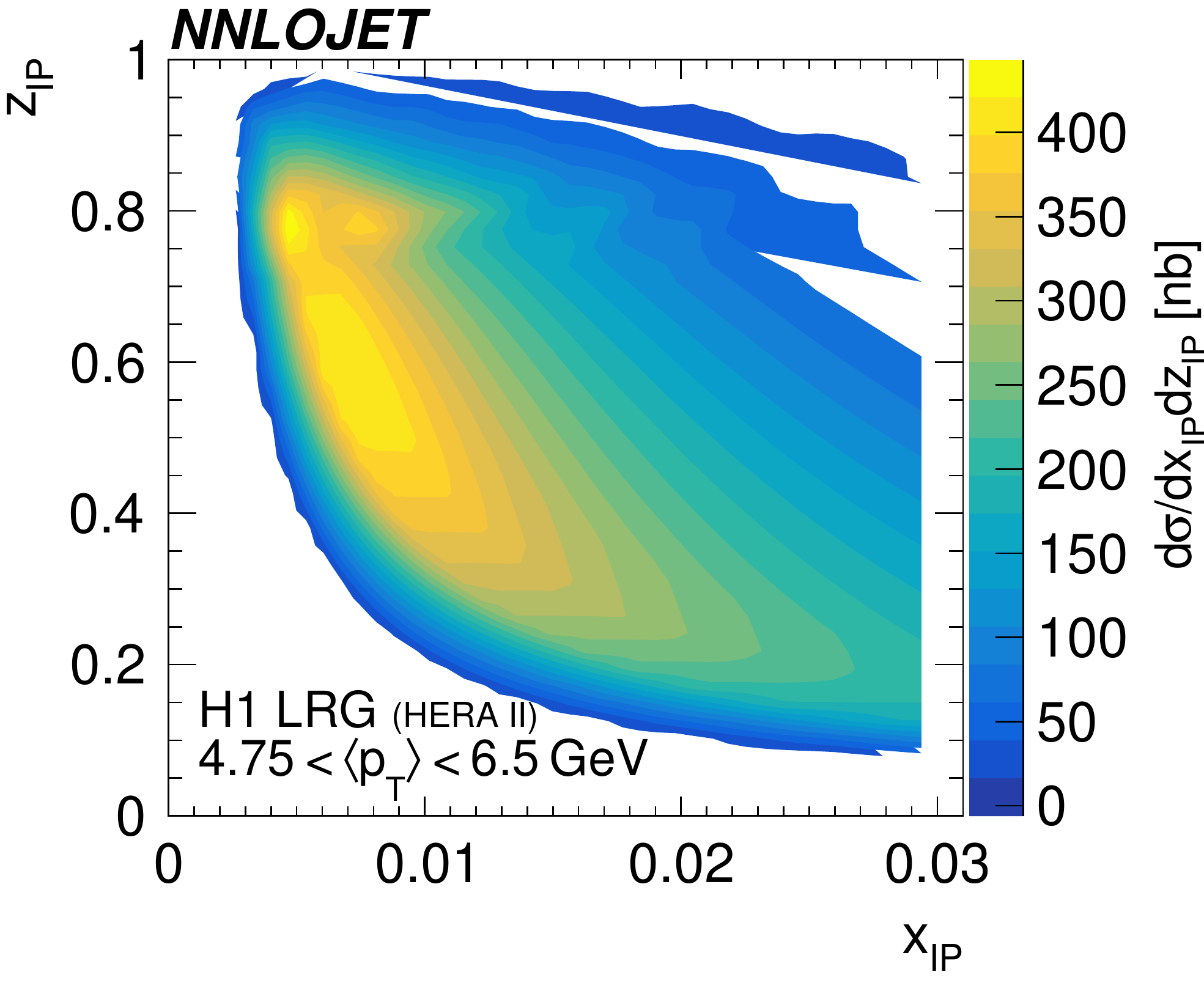} 
  \includegraphicss[width=0.32\textwidth]{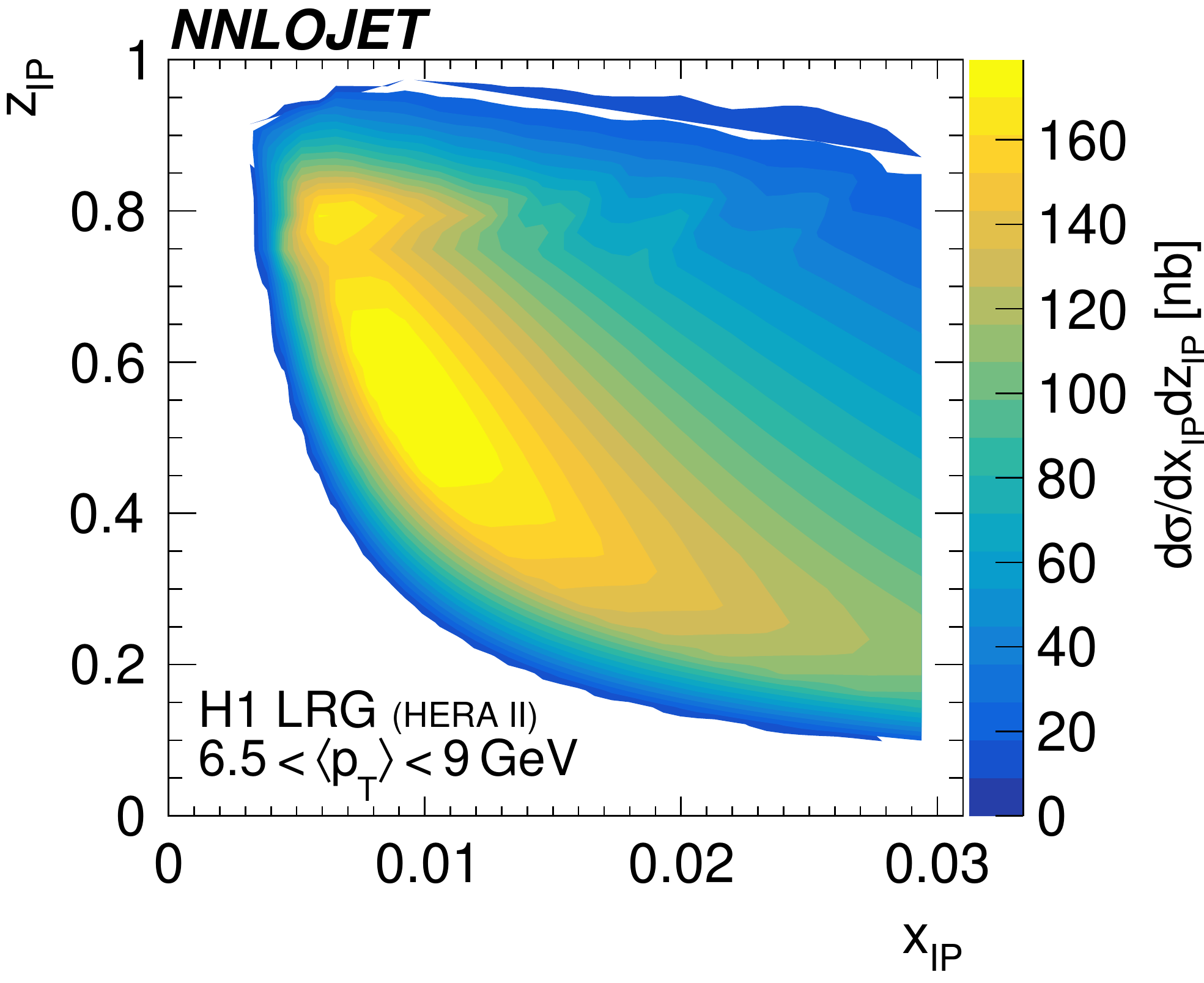}
  \includegraphicss[width=0.32\textwidth]{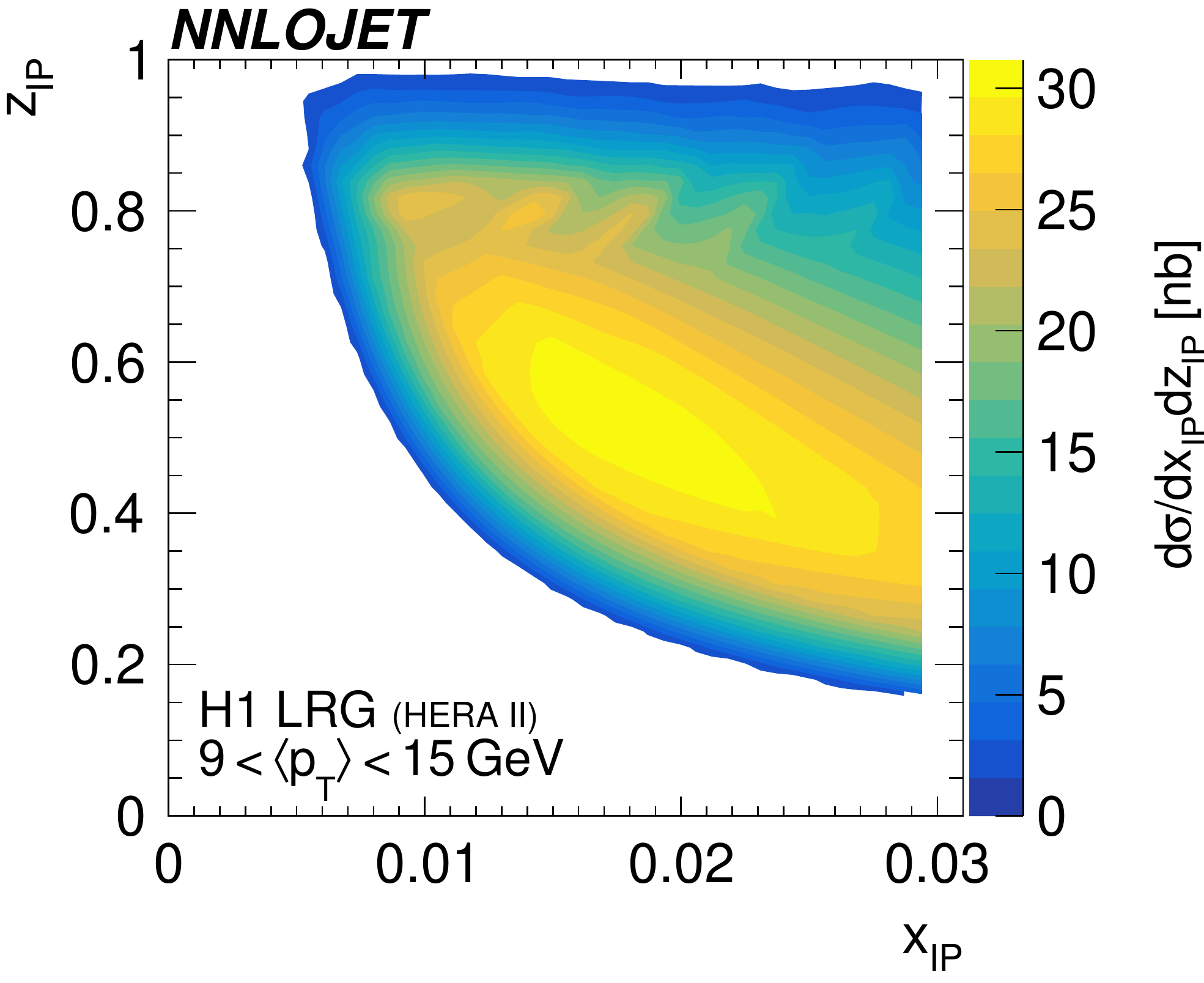}
  \caption{
    Contributions to the cross section for $\D\sigma/\D\ptavg$ of the
    \HLRG\ measurement as a function of \xpom and \zint\ (bin integrated).
    The three pads represent the three bins of this measurement.
    The color coding represents the differential
    cross sections as function of \zint\ and \xpom\ on a linear scale.
    The white areas are kinematically forbidden.
  }
\label{fig:xIPzIP}
\end{figure*}

A detailed study on the dependence of the cross section on the DPDF
is presented for the \ptavg\ distribution
of the \HLRG measurement.
The contributions to the cross section in each bin as a function of
the DPDF parameters \xpom and \zint\ is displayed in
figure~\ref{fig:xIPzIP}.
At highest values of \ptavg, only partons with comparably
high values of \xpom\ and \zint\ are contributing to the cross
section, whereas the cross section at medium values of \ptavg\ is
dominated by low \xpom\ and \zint\ partons.
All three bins have recognisable contributions from high values of \zpom
which is a distinct feature for predictions obtained with the \DPDFFitB DPDF.

\subsection{Quantitative comparison}
\begin{figure*}[tb]
  \centering
  \includegraphicss[trim={2.3cm 0 2.0cm 0},clip,width=0.98\textwidth]{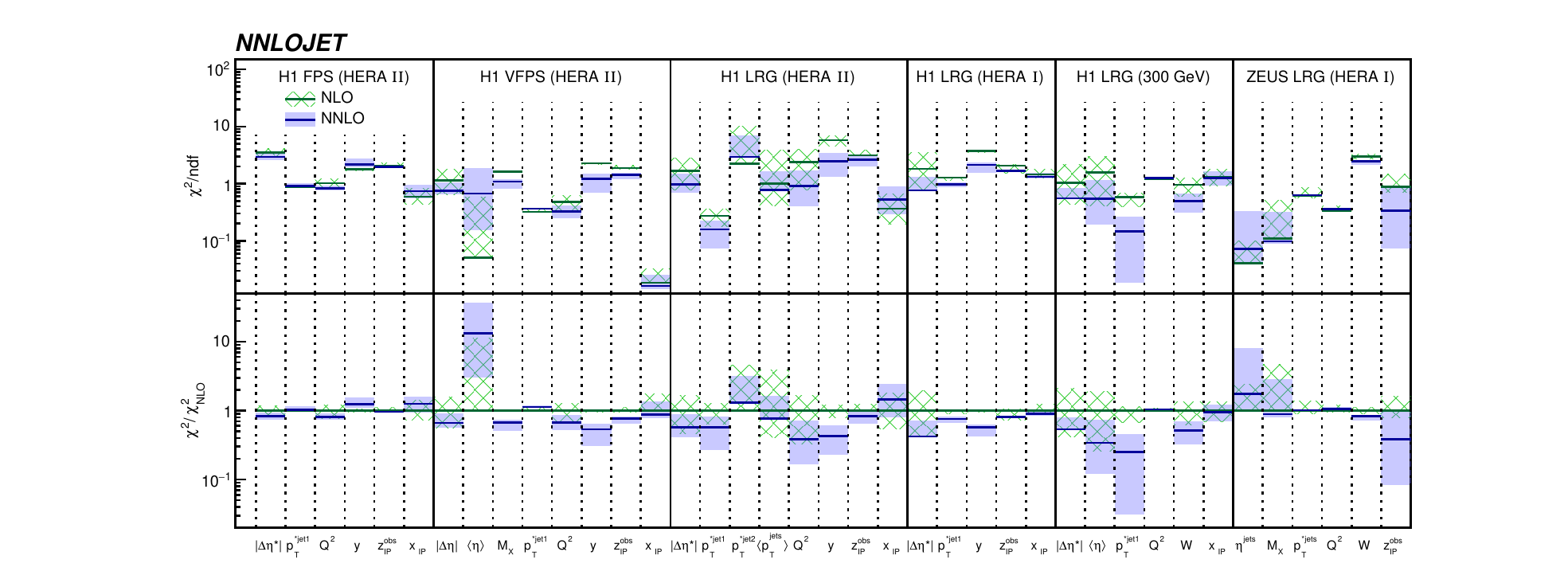} 
  \caption{The $\chisq/\ndf$ values for the
    analysed single-differential distributions obtained with NNLO and NLO predictions.
    The lower panel displays the ratio of $\chisq/\ndf$ to the NLO
    result.
    The size of the bands correspond to scale uncertainties.
    In all cases the \DPDFFitB DPDF was used.
   }
  \label{fig:chi2}
\end{figure*}

The agreement of NLO and NNLO predictions with data is quantified in
terms of a \chisq\ test. 
The \chisq\ function is defined as~\cite{Andreev:2014wwa}
\begin{equation}
  \chisq = 
  \sum_{i,j}
  \log\tfrac{\sigma_i^{\rm Data}}{\sigma_i^{\rm (N)NLO}}
  (V^{-1})_{ij}
  \log\tfrac{\sigma_j^{\rm Data}}{\sigma_j^{\rm (N)NLO}}\,,
  \label{eq:chi2}
\end{equation}
where the predictions $\sigma_{i,j}^{\rm (N)NLO}$ and data
$\sigma_{i,j}^{\rm data}$ for all points ($i$ or $j$) of a differential
distribution are considered and $V$ denotes the covariance matrix calculated from the relative experimental uncertainties.
We consider systematic uncertainties as fully correlated, if not
stated differently in the original publication.
In order to quantify only the agreement in shape, we consider the normalisation as
a free parameter and minimise \chisq\ with respect to it.
We calculate \chisq\ for all analysed single-differential
distributions.
Results for $\chisq/\ndf$ are displayed in figure~\ref{fig:chi2}.
For most of the distributions the $\chisq/\ndf$ values are smaller
when using NNLO rather than NLO predictions. 
The calculations are repeated for different DPDFs and different scale
functional forms and also in these cases, it is observed that
NNLO predictions mostly give lower $\chisq/\ndf$ values than 
NLO predictions (not shown).  
In an approximation, the normalisation of the predictions is
proportional to the gluon content of the DPDFs, whereas the shapes of
the differential distributions are related more closely to the hard matrix
elements.
Therefore, these results indicate that NNLO predictions provide a
better description of the data than NLO predictions, and we believe
that future DPDFs determined to NNLO QCD will be able to provide an improved
description of the dijet data, this also with respect to the normalisation.

From the double-differential distributions, we select the
$\D\sigma/\D\Qsq \D\ptjone$ measurement of the \HLRG\ analysis, and
data are compared to the NNLO and NLO predictions in
figure~\ref{fig:H1LRG2DptQ2}. 
For the \chisq\ evaluation, we minimise \chisq\ as a
function of \asmz, which is an equivalent procedure to the
\asmz\ determination presented previously by H1~\cite{Andreev:2014yra}.
The calculation using NLO predictions results in $\chisq/\ndf=16/14$.  
The calculation using NNLO predictions results in a value
of $\chisq/\ndf=13/14$, thus indicating also in this case an improved description
of the data. 
We estimate a scale uncertainty on the best fit value of \asmz\
with additional calculations using scale
factors of 0.5 and 2\,\footnote{The H1 
  collaboration estimated an uncertainty for \mur\ and
  \muf\ separately and considered the resulting uncertainties on the
  cross sections as half correlated and half uncorrelated.}.
The scale uncertainty of \asmz\ is found to be 11\,\% for the NLO
predictions, and for the NNLO predictions it is reduced to 4\,\%.
This reduction quantifies the significant improvement 
of the NNLO predictions as compared to NLO predictions.
The NNLO scale uncertainty is of similar size as the experimental one or
the DPDF uncertainties on \asmz, where H1 reported 4\,\% for
both~\cite{Andreev:2014yra}.
This study demonstrates that the NNLO calculations are suitable for further
phenomenological analyses, such as \asmz\ or DPDF fits, and the
NNLO scale uncertainties are of equal size as 
experimental uncertainties.

\section{Discussion and summary and conclusion}
\label{sec:summary}
We present the first NNLO QCD predictions for jet production in
diffractive scattering.  
Predictions for six measurements of dijet production in diffractive
deep-inelastic scattering from the H1 and ZEUS collaborations were
calculated and compared to data. 
We observe that the NNLO cross sections are significantly higher than
the data and are higher than NLO calculations by about 20\,\% to 40\,\% in
the studied kinematical range.

The NNLO predictions have significantly reduced scale uncertainties as
compared to NLO predictions.
The NNLO scale uncertainties are of similar size as the data
uncertainties and as the DPDF uncertainties.
Thus, the inclusion of dijet data in future DPDF analyses, and using NNLO
predictions, will allow for a substantial reduction of the uncertainty
of the gluon component.
The NNLO dijet calculations presented here are already in a numerical
format, which is suitable for such analyses.

Since no DPDFs in NNLO accuracy are available so far, only NLO DPDFs could be 
employed for our calculations.
The discrepancy of the NNLO predictions and data is believed to be due
to an overestimated gluon component of these DPDFs.
Alternative DPDFs, which also considered dijet data in their
determination, already result in typically lower NNLO predictions, but these still overshoot the data.
Ignoring the issue of normalisation, the shapes of differential
distributions are better described by NNLO than NLO predictions.
This is quantified by evaluating $\chisq$ values for the
examined experimental distributions. 
We believe that the normalisation difference between data and NNLO
predictions can be resolved by employing DPDFs determined to NNLO
accuracy and by including dijet data for their determinations. This in
particular as the gluon component is most important and is only weakly
 constrained by the inclusive data.

Furthermore, the comprehensive selection of all available dijet data represents 
the first comparison, where all these measurements are compared
to predictions obtained in an identical framework. 
Data taken with different experimental devices, at different
center-of-mass energies, and using either proton spectrometers or the
LRG method for the identification of the diffractive final state are
investigated.
All measurements are found to be mutually consistent when compared to
respective predictions.

The large amount of studied observables, which have so far not even been
studied in non-diffractive DIS, prove that NNLO predictions
provide an improved description of the data throughout.

The presented NNLO predictions provide an important step towards an
improved understanding of diffractive processes and represent a
precise test of the employed theoretical concepts.
In particular, it is observed, that for the given kinematical range of
the HERA data, higher-order corrections are of crucial importance,
while at the same time, no suitable DPDFs are currently available.

\begin{acknowledgements}
  We thank W.~Slominski and M.~Wing for discussions and help with the
  ZEUS data, and B.~Pokorny for discussions on the H1 LRG data.
  We thank X.~Chen, J.~Cruz-Martinez, R.~Gauld, A.~Gehrmann--De~Ridder,
  N.~Glover, M.~H{\"o}fer, I.~Majer, T.~Morgan, J.~Pires, D.~Walker
  and J.~Whitehead for useful discussions and their many contributions
  to the \NNLOJET\ code.  
  We are grateful for the collaboration with C.~Gwenlan, K.~Rabbertz and M.~Sutton
  for the interface of fastNLO and APPLgrid to \NNLOJET.
  This research was supported in part 
  by the Swiss National Science Foundation (SNF) under contract 200020-175595
  and  by the Research Executive Agency (REA) of the European Union under 
  the ERC Advanced Grant MC@NNLO (340983).
\end{acknowledgements}

\bibliographystyle{spphys}       
\bibliography{diffjets}   

\begin{thebibliography}{10}
\providecommand{\url}[1]{{#1}}
\providecommand{\urlprefix}{URL }
\expandafter\ifx\csname urlstyle\endcsname\relax
  \providecommand{\doi}[1]{DOI \discretionary{}{}{}#1}\else
  \providecommand{\doi}{DOI \discretionary{}{}{}\begingroup
  \urlstyle{rm}\Url}\fi

\bibitem{Newman:2013ada}
P.~Newman, M.~Wing, Rev. Mod. Phys. \textbf{86}(3), 1037 (2014).
\newblock \doi{10.1103/RevModPhys.86.1037}

\bibitem{Collins:1997sr}
J.C. Collins, Phys. Rev. \textbf{D57}, 3051 (1998).
\newblock \doi{10.1103/PhysRevD.61.019902, 10.1103/PhysRevD.57.3051}.
\newblock [Erratum: Phys. Rev.D 61, 019902 (2000)]

\bibitem{Andreev:2014yra}
V.~Andreev, et~al., JHEP \textbf{03}, 092 (2015).
\newblock \doi{10.1007/JHEP03(2015)092}

\bibitem{Ellis:1993tq}
S.D. Ellis, D.E. Soper, Phys. Rev. \textbf{D48}, 3160 (1993).
\newblock \doi{10.1103/PhysRevD.48.3160}

\bibitem{Altarelli:1977zs}
G.~Altarelli, G.~Parisi, Nucl. Phys. \textbf{B126}, 298 (1977).
\newblock \doi{10.1016/0550-3213(77)90384-4}

\bibitem{Gribov:1972ri}
V.N. Gribov, L.N. Lipatov, Sov. J. Nucl. Phys. \textbf{15}, 438 (1972).
\newblock [Yad.~Fiz.~15~(1972)~781]

\bibitem{Dokshitzer:1977sg}
Y.L. Dokshitzer, Sov. Phys. JETP \textbf{46}, 641 (1977).
\newblock [Zh.~Eksp.~Teor.~Fiz.~73~(1977)~1216]

\bibitem{Currie:2016ytq}
J.~Currie, T.~Gehrmann, J.~Niehues, Phys. Rev. Lett. \textbf{117}(4), 042001
  (2016).
\newblock \doi{10.1103/PhysRevLett.117.042001}

\bibitem{Currie:2017tpe}
J.~Currie, T.~Gehrmann, A.~Huss, J.~Niehues, JHEP \textbf{07}, 018 (2017).
\newblock \doi{10.1007/JHEP07(2017)018}

\bibitem{Garland:2001tf}
L.W. Garland, T.~Gehrmann, E.W.N. Glover, A.~Koukoutsakis, E.~Remiddi, Nucl.
  Phys. \textbf{B627}, 107 (2002).
\newblock \doi{10.1016/S0550-3213(02)00057-3}

\bibitem{Garland:2002ak}
L.W. Garland, T.~Gehrmann, E.W.N. Glover, A.~Koukoutsakis, E.~Remiddi, Nucl.
  Phys. \textbf{B642}, 227 (2002).
\newblock \doi{10.1016/S0550-3213(02)00627-2}

\bibitem{Gehrmann:2002zr}
T.~Gehrmann, E.~Remiddi, Nucl. Phys. \textbf{B640}, 379 (2002).
\newblock \doi{10.1016/S0550-3213(02)00569-2}

\bibitem{Gehrmann:2009vu}
T.~Gehrmann, E.W.N. Glover, Phys. Lett. \textbf{B676}, 146 (2009).
\newblock \doi{10.1016/j.physletb.2009.04.083}

\bibitem{Glover:1996eh}
E.W.N. Glover, D.J. Miller, Phys. Lett. \textbf{B396}, 257 (1997).
\newblock \doi{10.1016/S0370-2693(97)00113-5}

\bibitem{Bern:1996ka}
Z.~Bern, L.J. Dixon, D.A. Kosower, S.~Weinzierl, Nucl. Phys. \textbf{B489}, 3
  (1997).
\newblock \doi{10.1016/S0550-3213(96)00703-1}

\bibitem{Campbell:1997tv}
J.M. Campbell, E.W.N. Glover, D.J. Miller, Phys. Lett. \textbf{B409}, 503
  (1997).
\newblock \doi{10.1016/S0370-2693(97)00909-X}

\bibitem{Bern:1997sc}
Z.~Bern, L.J. Dixon, D.A. Kosower, Nucl. Phys. \textbf{B513}, 3 (1998).
\newblock \doi{10.1016/S0550-3213(97)00703-7}

\bibitem{Hagiwara:1988pp}
K.~Hagiwara, D.~Zeppenfeld, Nucl. Phys. \textbf{B313}, 560 (1989).
\newblock \doi{10.1016/0550-3213(89)90397-0}

\bibitem{Berends:1988yn}
F.A. Berends, W.T. Giele, H.~Kuijf, Nucl. Phys. \textbf{B321}, 39 (1989).
\newblock \doi{10.1016/0550-3213(89)90242-3}

\bibitem{Falck:1989uz}
N.K. Falck, D.~Graudenz, G.~Kramer, Nucl. Phys. \textbf{B328}, 317 (1989).
\newblock \doi{10.1016/0550-3213(89)90331-3}

\bibitem{Sterman:1977wj}
G.F. Sterman, S.~Weinberg, Phys. Rev. Lett. \textbf{39}, 1436 (1977).
\newblock \doi{10.1103/PhysRevLett.39.1436}

\bibitem{GehrmannDeRidder:2005cm}
A.~Gehrmann-De~Ridder, T.~Gehrmann, E.W.N. Glover, JHEP \textbf{09}, 056
  (2005).
\newblock \doi{10.1088/1126-6708/2005/09/056}

\bibitem{GehrmannDeRidder:2005aw}
A.~Gehrmann-De~Ridder, T.~Gehrmann, E.W.N. Glover, Phys. Lett. \textbf{B612},
  49 (2005).
\newblock \doi{10.1016/j.physletb.2005.03.003}

\bibitem{GehrmannDeRidder:2005hi}
A.~Gehrmann-De~Ridder, T.~Gehrmann, E.W.N. Glover, Phys. Lett. \textbf{B612},
  36 (2005).
\newblock \doi{10.1016/j.physletb.2005.02.039}

\bibitem{Currie:2013vh}
J.~Currie, E.W.N. Glover, S.~Wells, JHEP \textbf{04}, 066 (2013).
\newblock \doi{10.1007/JHEP04(2013)066}

\bibitem{Gleisberg:2008ta}
T.~Gleisberg, S.~Hoeche, F.~Krauss, M.~Schonherr, S.~Schumann, F.~Siegert,
  J.~Winter, JHEP \textbf{02}, 007 (2009).
\newblock \doi{10.1088/1126-6708/2009/02/007}

\bibitem{Carli:2010cg}
T.~Carli, T.~Gehrmann, S.~Hoeche, Eur. Phys. J. \textbf{C67}, 73 (2010).
\newblock \doi{10.1140/epjc/s10052-010-1261-2}

\bibitem{Cascioli:2011va}
F.~Cascioli, P.~Maierhofer, S.~Pozzorini, Phys. Rev. Lett. \textbf{108}, 111601
  (2012).
\newblock \doi{10.1103/PhysRevLett.108.111601}

\bibitem{Nagy:2001xb}
Z.~Nagy, Z.~Trocsanyi, Phys. Rev. Lett. \textbf{87}, 082001 (2001).
\newblock \doi{10.1103/PhysRevLett.87.082001}

\bibitem{Nagy:2003tz}
Z.~Nagy, Phys. Rev. \textbf{D68}, 094002 (2003).
\newblock \doi{10.1103/PhysRevD.68.094002}

\bibitem{Gehrmann:2018szu}
T.~Gehrmann, et~al., in \emph{{Proceedings, 13th International Symposium on
  Radiative Corrections: Application of Quantum Field Theory to Phenomenology
  (RADCOR2017): St. Gilgen, Austria, September 24-29, 2017}} (2018).
\newblock
  \urlprefix\url{http://inspirehep.net/record/1649093/files/arXiv:1801.06415.pdf}

\bibitem{Glover:2010im}
E.W.N. Glover, J.~Pires, JHEP \textbf{06}, 096 (2010).
\newblock \doi{10.1007/JHEP06(2010)096}

\bibitem{GehrmannDeRidder:2011aa}
A.~Gehrmann-De~Ridder, E.W.N. Glover, J.~Pires, JHEP \textbf{02}, 141 (2012).
\newblock \doi{10.1007/JHEP02(2012)141}

\bibitem{Patrignani:2016xqp}
C.~Patrignani, et~al., Chin. Phys. \textbf{C40}(10), 100001 (2016).
\newblock \doi{10.1088/1674-1137/40/10/100001}

\bibitem{Andreev:2016tgi}
V.~Andreev, et~al., Eur. Phys. J. \textbf{C77}(4), 215 (2017).
\newblock \doi{10.1140/epjc/s10052-017-4717-9}

\bibitem{Andreev:2017vxu}
V.~Andreev, et~al., Eur. Phys. J. \textbf{C77}(11), 791 (2017).
\newblock \doi{10.1140/epjc/s10052-017-5314-7}

\bibitem{Aktas:2007bv}
A.~Aktas, et~al., JHEP \textbf{10}, 042 (2007).
\newblock \doi{10.1088/1126-6708/2007/10/042}

\bibitem{Britzger:2012bs}
D.~Britzger, K.~Rabbertz, F.~Stober, M.~Wobisch, in \emph{{Proceedings, 20th
  International Workshop on Deep-Inelastic Scattering and Related Subjects (DIS
  2012): Bonn, Germany, March 26-30, 2012}} (2012), p. 217.
\newblock \doi{10.3204/DESY-PROC-2012-02/165}.
\newblock
  \urlprefix\url{http://inspirehep.net/record/1128033/files/arXiv:1208.3641.pdf}

\bibitem{Britzger:2013kia}
D.~Britzger, {Regularized Unfolding of Jet Cross Sections in Deep-Inelastic
  $ep$ Scattering at HERA and Determination of the Strong Coupling Constant}.
\newblock Ph.D. thesis, U. Hamburg, Dept. Phys. (2013).
\newblock \doi{10.3204/DESY-THESIS-2013-045}.
\newblock
  \urlprefix\url{http://www-library.desy.de/cgi-bin/showprep.pl?thesis13-045}

\bibitem{Potter:1999gg}
B.~Potter, Comput. Phys. Commun. \textbf{133}, 105 (2000).
\newblock \doi{10.1016/S0010-4655(00)00158-2}

\bibitem{Duprel:1999wz}
C.~Duprel, T.~Hadig, N.~Kauer, M.~Wobisch, in \emph{{Monte Carlo generators for
  HERA physics. Proceedings, Workshop, Hamburg, Germany, 1998-1999}} (1999), p.
  142

\bibitem{Ingelman:1984ns}
G.~Ingelman, P.E. Schlein, Phys. Lett. \textbf{152B}, 256 (1985).
\newblock \doi{10.1016/0370-2693(85)91181-5}

\bibitem{Ahmed:1995ns}
T.~Ahmed, et~al., Phys. Lett. \textbf{B348}, 681 (1995).
\newblock \doi{10.1016/0370-2693(95)00279-T}

\bibitem{Gehrmann:1995by}
T.~Gehrmann, W.J. Stirling, Z. Phys. \textbf{C70}, 89 (1996).
\newblock \doi{10.1007/s002880050085}

\bibitem{Aaron:2012ad}
F.D. Aaron, et~al., Eur. Phys. J. \textbf{C72}, 2074 (2012).
\newblock \doi{10.1140/epjc/s10052-012-2074-2}

\bibitem{Aktas:2006hy}
A.~Aktas, et~al., Eur. Phys. J. \textbf{C48}, 715 (2006).
\newblock \doi{10.1140/epjc/s10052-006-0035-3}

\bibitem{Chekanov:2009aa}
S.~Chekanov, et~al., Nucl. Phys. \textbf{B831}, 1 (2010).
\newblock \doi{10.1016/j.nuclphysb.2010.01.014}

\bibitem{Martin:2006td}
A.D. Martin, M.G. Ryskin, G.~Watt, Phys. Lett. \textbf{B644}, 131 (2007).
\newblock \doi{10.1016/j.physletb.2006.11.032}

\bibitem{Goharipour:2018yov}
M.~Goharipour, H.~Khanpour, V.~Guzey,   (2018)

\bibitem{Aaron:2010aa}
F.D. Aaron, et~al., Eur. Phys. J. \textbf{C71}, 1578 (2011).
\newblock \doi{10.1140/epjc/s10052-011-1578-5}

\bibitem{Aktas:2006hx}
A.~Aktas, et~al., Eur. Phys. J. \textbf{C48}, 749 (2006).
\newblock \doi{10.1140/epjc/s10052-006-0046-0}

\bibitem{List:1998jz}
B.~List, A.~Mastroberardino, Conf. Proc. \textbf{C980427}, 396 (1998)

\bibitem{Aaron:2011mp}
F.D. Aaron, et~al., Eur. Phys. J. \textbf{C72}, 1970 (2012).
\newblock \doi{10.1140/epjc/s10052-012-1970-9}

\bibitem{Andreev:2015cwa}
V.~Andreev, et~al., JHEP \textbf{05}, 056 (2015).
\newblock \doi{10.1007/JHEP05(2015)056}

\bibitem{Aktas:2007hn}
A.~Aktas, et~al., Eur. Phys. J. \textbf{C51}, 549 (2007).
\newblock \doi{10.1140/epjc/s10052-007-0325-4}

\bibitem{Chekanov:2007aa}
S.~Chekanov, et~al., Eur. Phys. J. \textbf{C52}, 813 (2007).
\newblock \doi{10.1140/epjc/s10052-007-0426-0, 10.3204/proc07-01/112}.
\newblock [,671(2007)]

\bibitem{Bonato:2008zz}
A.~Bonato, {Diffractive Dijet Production in Deep Inelastic Scattering at ZEUS}.
\newblock Ph.D. thesis, Hamburg U. (2008).
\newblock \doi{10.3204/DESY-THESIS-2008-008}.
\newblock
  \urlprefix\url{http://inspirehep.net/record/783511/files/desy-thesis-08-008.pdf}

\bibitem{Zlebcik:2011kq}
R.~Zlebcik, K.~Cerny, A.~Valkarova, Eur. Phys. J. \textbf{C71}, 1741 (2011).
\newblock \doi{10.1140/epjc/s10052-011-1741-z}

\bibitem{Andreev:2014wwa}
V.~Andreev, et~al., Eur. Phys. J. C \textbf{75}, 65 (2015).
\newblock \doi{10.1140/epjc/s10052-014-3223-6}

\end{thebibliography}


\end{document}